\documentclass[amsmath,amssymb,aps,prd,11pt,tightenlines,superscriptaddress,nofootinbib,preprintnumbers,notitlepage]{revtex4-2}

\usepackage{amsmath,amssymb,amsthm,amsfonts}
\usepackage{graphicx,tabularx}
\usepackage{color}
\usepackage{multirow}
\usepackage{comment}
\usepackage{enumitem}
\usepackage{subfigure}
\usepackage{orcidlink}
\usepackage{cleveref}
\usepackage{slashed}
\usepackage[normalem]{ulem}
\usepackage{scalerel}
\usepackage{wrapfig}
\usepackage{cancel}
\usepackage{xcolor}
\usepackage{multirow}
\usepackage{setspace}
\usepackage{lineno}

\newcommand{\PRE}[1]{{#1}} 

\newcommand{\be}{\begin{equation} \begin{aligned}}
\newcommand{\ee}{\end{aligned} \end{equation}}
\newcommand{\beqa}{\begin{eqnarray}}
\newcommand{\eeqa}{\end{eqnarray}}

\def\figureautorefname~#1\null{Fig.\,#1\null}
\def\tableautorefname~#1\null{Tab.\,#1\null}
\def\equationautorefname~#1\null{Eq.\,(#1)\null}

\newcommand{\iab}{\text{ab}^{-1}}

\newcommand{\ifb}{\text{fb}^{-1}}

\newcommand{\cm}{\text{cm}}
\newcommand{\m}{\text{m}}

\RequirePackage[normalem]{ulem}

\crefname{section}{Sec.}{Secs.}
\crefname{figure}{Fig.}{Figs.}
\crefname{equation}{Eq.}{Eqs.}
\crefname{table}{Table}{Tables}
\crefname{appendix}{Appendix}{Appendices}

\setlist[enumerate]{leftmargin=0.5cm}

\makeatletter
\renewcommand{\p@subsection}{}
\renewcommand{\p@subsubsection}{}
\makeatother

\usepackage{xspace}
\newcommand{\fasernu}{FASER$\nu$\xspace}

\newcommand{\figref}[1]{Fig.~\ref{fig:#1}}

\newcommand{\myaffiliation}[1]{\affiliation{#1}}

\begin{document}

\preprint{CERN-FASER-2025-001}

\title{{\Large Prospects and Opportunities with an upgraded FASER Neutrino Detector during the HL-LHC era: Input to the EPPSU} 
\medskip \\  
FASER Collaboration
}

\begin{figure*}[h]
\vspace*{-0.6in}
\begin{flushleft}
\includegraphics[width=0.19\textwidth]{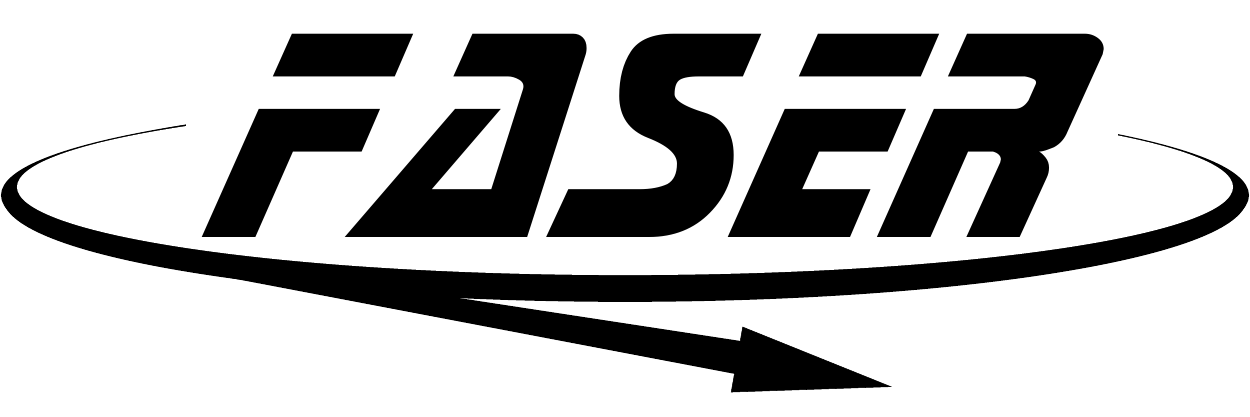}
\end{flushleft}
\end{figure*}

\author{Roshan Mammen Abraham\,\orcidlink{0000-0003-4678-3808}}
\myaffiliation{Department of Physics and Astronomy, University of California, Irvine, CA 92697-4575, USA}

\author{Xiaocong Ai\,\orcidlink{0000-0003-3856-2415}}
\myaffiliation{School of Physics, Zhengzhou University, Zhengzhou 450001, China}

\author{Saul Alonso-Monsalve\,\orcidlink{0000-0002-9678-7121}}
\myaffiliation{Institute for Particle Physics, ETH Zurich, 8093 Zurich, Switzerland}

\author{John Anders\,\orcidlink{0000-0002-1846-0262}}
\myaffiliation{University of Liverpool, Liverpool L69 3BX, United Kingdom}

\author{Claire Antel\,\orcidlink{0000-0001-9683-0890}}
\myaffiliation{D\'epartement de Physique Nucl\'eaire et Corpusculaire, University of Geneva, CH-1211 Geneva 4, Switzerland}

\author{Akitaka Ariga\,\orcidlink{0000-0002-6832-2466}}
\myaffiliation{Albert Einstein Center for Fundamental Physics, Laboratory for High Energy Physics, University of Bern, Sidlerstrasse 5, CH-3012 Bern, Switzerland}
\myaffiliation{Department of Physics, Chiba University, 1-33 Yayoi-cho Inage-ku, 263-8522 Chiba, Japan}

\author{Tomoko Ariga\,\orcidlink{0000-0001-9880-3562}}
\myaffiliation{Kyushu University, 744 Motooka, Nishi-ku, 819-0395 Fukuoka, Japan }

\author{Jeremy Atkinson\,\orcidlink{0009-0003-3287-2196}}
\myaffiliation{Albert Einstein Center for Fundamental Physics, Laboratory for High Energy Physics, University of Bern, Sidlerstrasse 5, CH-3012 Bern, Switzerland}

\author{Florian~U.~Bernlochner\,\orcidlink{0000-0001-8153-2719}}
\myaffiliation{Universit\"at Bonn, Regina-Pacis-Weg 3, D-53113 Bonn, Germany}

\author{Tobias Boeckh\,\orcidlink{0009-0000-7721-2114}}
\myaffiliation{Universit\"at Bonn, Regina-Pacis-Weg 3, D-53113 Bonn, Germany}

\author{Jamie Boyd\,\orcidlink{0000-0001-7360-0726}}
\myaffiliation{CERN, CH-1211 Geneva 23, Switzerland}

\author{Lydia Brenner\,\orcidlink{0000-0001-5350-7081}}
\myaffiliation{Nikhef National Institute for Subatomic Physics, Science Park 105, 1098 XG Amsterdam, Netherlands}

\author{Angela Burger\,\orcidlink{0000-0003-0685-4122}}
\myaffiliation{CERN, CH-1211 Geneva 23, Switzerland}

\author{Franck Cadoux} 
\myaffiliation{D\'epartement de Physique Nucl\'eaire et Corpusculaire, University of Geneva, CH-1211 Geneva 4, Switzerland}

\author{Roberto Cardella\,\orcidlink{0000-0002-3117-7277}}
\myaffiliation{D\'epartement de Physique Nucl\'eaire et Corpusculaire, University of Geneva, CH-1211 Geneva 4, Switzerland}

\author{David~W.~Casper\,\orcidlink{0000-0002-7618-1683}}
\myaffiliation{Department of Physics and Astronomy, University of California, Irvine, CA 92697-4575, USA}

\author{Charlotte Cavanagh\,\orcidlink{0009-0001-1146-5247}}
\myaffiliation{Institute for Particle Physics, ETH Zurich, 8093 Zurich, Switzerland}

\author{Xin Chen\,\orcidlink{0000-0003-4027-3305}}
\myaffiliation{Department of Physics, Tsinghua University, Beijing, China}

\author{Dhruv Chouhan\,\orcidlink{0009-0007-2664-0742}}
\myaffiliation{Universit\"at Bonn, Regina-Pacis-Weg 3, D-53113 Bonn, Germany}

\author{Sebastiani Christiano\,\orcidlink{0000-0003-1073-035X}}
\myaffiliation{CERN, CH-1211 Geneva 23, Switzerland}

\author{Andrea Coccaro\,\orcidlink{0000-0003-2368-4559}}
\myaffiliation{INFN Sezione di Genova, Via Dodecaneso, 33--16146, Genova, Italy}

\author{Stephane D\'{e}bieux} 
\myaffiliation{D\'epartement de Physique Nucl\'eaire et Corpusculaire, University of Geneva, CH-1211 Geneva 4, Switzerland}

\author{Monica D’Onofrio\,\orcidlink{0000-0003-2408-5099}}
\myaffiliation{University of Liverpool, Liverpool L69 3BX, United Kingdom}

\author{Ansh Desai\,\orcidlink{0000-0002-5447-8304}}
\myaffiliation{University of Oregon, Eugene, OR 97403, USA}

\author{Sergey Dmitrievsky\,\orcidlink{0000-0003-4247-8697}}
\myaffiliation{Affiliated with an international laboratory covered by a cooperation agreement with CERN.}

\author{Radu Dobre\,\orcidlink{0000-0002-9518-6068}}
\myaffiliation{Institute of Space Science---INFLPR Subsidiary, Bucharest, Romania}

\author{Sinead Eley\,\orcidlink{0009-0001-1320-2889}}
\myaffiliation{University of Liverpool, Liverpool L69 3BX, United Kingdom}

\author{Yannick Favre} 
\myaffiliation{D\'epartement de Physique Nucl\'eaire et Corpusculaire, University of Geneva, CH-1211 Geneva 4, Switzerland}

\author{Jonathan~L.~Feng\,\orcidlink{0000-0002-7713-2138}}
\myaffiliation{Department of Physics and Astronomy, University of California, Irvine, CA 92697-4575, USA}

\author{Carlo Alberto Fenoglio\,\orcidlink{0009-0007-7567-8763}}
\myaffiliation{D\'epartement de Physique Nucl\'eaire et Corpusculaire, University of Geneva, CH-1211 Geneva 4, Switzerland}

\author{Didier Ferrere\,\orcidlink{0000-0002-5687-9240}}
\myaffiliation{D\'epartement de Physique Nucl\'eaire et Corpusculaire, University of Geneva, CH-1211 Geneva 4, Switzerland}

\author{Max Fieg\,\orcidlink{0000-0002-7027-6921}}
\myaffiliation{Department of Physics and Astronomy, University of California, Irvine, CA 92697-4575, USA}

\author{Wissal Filali\,\orcidlink{0009-0008-6961-2335}}
\myaffiliation{Universit\"at Bonn, Regina-Pacis-Weg 3, D-53113 Bonn, Germany}

\author{Elena Firu\,\orcidlink{0000-0002-3109-5378}}
\myaffiliation{Institute of Space Science---INFLPR Subsidiary, Bucharest, Romania}

\author{Edward Galantay\,\orcidlink{0009-0001-6749-7360}}
\myaffiliation{CERN, CH-1211 Geneva 23, Switzerland}
\myaffiliation{D\'epartement de Physique Nucl\'eaire et Corpusculaire, University of Geneva, CH-1211 Geneva 4, Switzerland}

\author{Ali Garabaglu\,\orcidlink{0000-0002-8105-6027}}
\myaffiliation{Department of Physics, University of Washington, PO Box 351560, Seattle, WA 98195-1460, USA}

\author{Stephen Gibson\,\orcidlink{0000-0002-1236-9249}}
\myaffiliation{Royal Holloway, University of London, Egham, TW20 0EX, United Kingdom}

\author{Sergio Gonzalez-Sevilla\,\orcidlink{0000-0003-4458-9403}}
\myaffiliation{D\'epartement de Physique Nucl\'eaire et Corpusculaire, University of Geneva, CH-1211 Geneva 4, Switzerland}

\author{Yuri Gornushkin\,\orcidlink{0000-0003-3524-4032}}
\myaffiliation{Affiliated with an international laboratory covered by a cooperation agreement with CERN.}

\author{Carl Gwilliam\,\orcidlink{0000-0002-9401-5304}}
\myaffiliation{University of Liverpool, Liverpool L69 3BX, United Kingdom}

\author{Daiki Hayakawa\,\orcidlink{0000-0003-4253-4484}}
\myaffiliation{Department of Physics, Chiba University, 1-33 Yayoi-cho Inage-ku, 263-8522 Chiba, Japan}

\author{Michael Holzbock\,\orcidlink{0000-0001-8018-4185}}
\myaffiliation{CERN, CH-1211 Geneva 23, Switzerland}

\author{Shih-Chieh Hsu\,\orcidlink{0000-0001-6214-8500}}
\myaffiliation{Department of Physics, University of Washington, PO Box 351560, Seattle, WA 98195-1460, USA}

\author{Zhen Hu\,\orcidlink{0000-0001-8209-4343}}
\myaffiliation{Department of Physics, Tsinghua University, Beijing, China}

\author{Giuseppe Iacobucci\,\orcidlink{0000-0001-9965-5442}}
\myaffiliation{D\'epartement de Physique Nucl\'eaire et Corpusculaire, University of Geneva, CH-1211 Geneva 4, Switzerland}

\author{Tomohiro Inada\,\orcidlink{0000-0002-6923-9314}}
\myaffiliation{Kyushu University, 744 Motooka, Nishi-ku, 819-0395 Fukuoka, Japan }

\author{Luca Iodice\,\orcidlink{0000-0002-3516-7121}}
\myaffiliation{D\'epartement de Physique Nucl\'eaire et Corpusculaire, University of Geneva, CH-1211 Geneva 4, Switzerland}

\author{Sune Jakobsen\,\orcidlink{0000-0002-6564-040X}}
\myaffiliation{CERN, CH-1211 Geneva 23, Switzerland}

\author{Hans Joos\,\orcidlink{0000-0003-4313-4255}}
\myaffiliation{CERN, CH-1211 Geneva 23, Switzerland}
\myaffiliation{II.~Physikalisches Institut, Universität Göttingen, Göttingen, Germany}

\author{Enrique Kajomovitz\,\orcidlink{0000-0002-8464-1790}}
\myaffiliation{Department of Physics and Astronomy, Technion---Israel Institute of Technology, Haifa 32000, Israel}

\author{Hiroaki Kawahara\,\orcidlink{0009-0007-5657-9954}}
\myaffiliation{Kyushu University, 744 Motooka, Nishi-ku, 819-0395 Fukuoka, Japan}

\author{Alex Keyken\,\orcidlink{0009-0001-4886-2924}}
\myaffiliation{Royal Holloway, University of London, Egham, TW20 0EX, United Kingdom}

\author{Felix Kling\,\orcidlink{0000-0002-3100-6144}}
\myaffiliation{Deutsches Elektronen-Synchrotron DESY, Notkestr.~85, 22607 Hamburg, Germany}

\author{Daniela Köck\,\orcidlink{0000-0002-9090-5502}}
\myaffiliation{University of Oregon, Eugene, OR 97403, USA}

\author{Pantelis Kontaxakis\,\orcidlink{0000-0002-4860-5979}}
\myaffiliation{D\'epartement de Physique Nucl\'eaire et Corpusculaire, University of Geneva, CH-1211 Geneva 4, Switzerland}

\author{Umut Kose\,\orcidlink{0000-0001-5380-9354}}
\myaffiliation{Institute for Particle Physics, ETH Zurich, 8093 Zurich, Switzerland}

\author{Rafaella Kotitsa\,\orcidlink{0000-0002-7886-2685}}
\myaffiliation{CERN, CH-1211 Geneva 23, Switzerland}

\author{Peter Krack\,\orcidlink{0009-0003-5694-887X}}
\myaffiliation{Nikhef National Institute for Subatomic Physics, Science Park 105, 1098 XG Amsterdam, Netherlands}

\author{Susanne Kuehn\,\orcidlink{0000-0001-5270-0920}}
\myaffiliation{CERN, CH-1211 Geneva 23, Switzerland}

\author{Thanushan Kugathasan\,\orcidlink{0000-0003-4631-5019}}
\myaffiliation{D\'epartement de Physique Nucl\'eaire et Corpusculaire, University of Geneva, CH-1211 Geneva 4, Switzerland}

\author{Lorne Levinson\,\orcidlink{0000-0003-4679-0485}}
\myaffiliation{Department of Particle Physics and Astrophysics, Weizmann Institute of Science, Rehovot 76100, Israel}

\author{Botao Li\,\orcidlink{0009-0009-0097-3367}}
\myaffiliation{Institute for Particle Physics, ETH Zurich, 8093 Zurich, Switzerland}

\author{Jinfeng Liu\,\orcidlink{0000-0001-6827-1729}}
\myaffiliation{Department of Physics, Tsinghua University, Beijing, China}

\author{Yi Liu\,\orcidlink{0000-0002-3576-7004}}
\myaffiliation{School of Physics, Zhengzhou University, Zhengzhou 450001, China}

\author{Margaret~S.~Lutz\,\orcidlink{0000-0003-4515-0224}}
\myaffiliation{CERN, CH-1211 Geneva 23, Switzerland}

\author{Jack MacDonald\,\orcidlink{0000-0002-3150-3124}}
\myaffiliation{Institut f\"ur Physik, Universität Mainz, Mainz, Germany}

\author{Chiara Magliocca\,\orcidlink{0009-0009-4927-9253}}
\myaffiliation{D\'epartement de Physique Nucl\'eaire et Corpusculaire, University of Geneva, CH-1211 Geneva 4, Switzerland}

\author{Toni~M\"akel\"a\,\orcidlink{0000-0002-1723-4028}}
\myaffiliation{Department of Physics and Astronomy, University of California, Irvine, CA 92697-4575, USA}

\author{Yasuhiro Maruya\,\orcidlink{0009-0008-5349-176X}}
\myaffiliation{Institute of Science Tokyo, 2-12-1, Ookayama, Meguro-ku, 152-8550 Tokyo, Japan}

\author{Lawson McCoy\,\orcidlink{0009-0009-2741-3220}}
\myaffiliation{Department of Physics and Astronomy, University of California, Irvine, CA 92697-4575, USA}

\author{Josh McFayden\,\orcidlink{0000-0001-9273-2564}}
\myaffiliation{Department of Physics \& Astronomy, University of Sussex, Sussex House, Falmer, Brighton, BN1 9RH, United Kingdom}

\author{Andrea Pizarro Medina\,\orcidlink{0000-0002-1024-5605}}
\myaffiliation{D\'epartement de Physique Nucl\'eaire et Corpusculaire, University of Geneva, CH-1211 Geneva 4, Switzerland}

\author{Matteo Milanesio\,\orcidlink{0000-0001-8778-9638}}
\myaffiliation{D\'epartement de Physique Nucl\'eaire et Corpusculaire, University of Geneva, CH-1211 Geneva 4, Switzerland}

\author{Théo Moretti\,\orcidlink{0000-0001-7065-1923}}
\myaffiliation{D\'epartement de Physique Nucl\'eaire et Corpusculaire, University of Geneva, CH-1211 Geneva 4, Switzerland}

\author{Mitsuhiro Nakamura\,\orcidlink{0009-0002-6032-2741}}
\myaffiliation{Nagoya University, Furo-cho, Chikusa-ku, Nagoya 464-8602, Japan}

\author{Toshiyuki Nakano\,\orcidlink{0009-0004-8568-9077}}
\myaffiliation{Nagoya University, Furo-cho, Chikusa-ku, Nagoya 464-8602, Japan}

\author{Laurie Nevay\,\orcidlink{0000-0001-7225-9327}}
\myaffiliation{CERN, CH-1211 Geneva 23, Switzerland}

\author{Ken Ohashi\,\orcidlink{0009-0000-9494-8457}}
\myaffiliation{Albert Einstein Center for Fundamental Physics, Laboratory for High Energy Physics, University of Bern, Sidlerstrasse 5, CH-3012 Bern, Switzerland}

\author{Hidetoshi Otono\,\orcidlink{0000-0003-0760-5988}}
\myaffiliation{Kyushu University, 744 Motooka, Nishi-ku, 819-0395 Fukuoka, Japan}

\author{Hao Pang\,\orcidlink{0000-0002-1946-1769}}
\myaffiliation{Department of Physics, Tsinghua University, Beijing, China}

\author{Lorenzo Paolozzi\,\orcidlink{0000-0002-9281-1972}}
\myaffiliation{D\'epartement de Physique Nucl\'eaire et Corpusculaire, University of Geneva, CH-1211 Geneva 4, Switzerland}
\myaffiliation{CERN, CH-1211 Geneva 23, Switzerland}

\author{Pawan Pawan\,\orcidlink{0009-0004-9339-5984}}
\myaffiliation{University of Liverpool, Liverpool L69 3BX, United Kingdom}

\author{Brian Petersen\,\orcidlink{0000-0002-7380-6123}}
\myaffiliation{CERN, CH-1211 Geneva 23, Switzerland}

\author{Titi Preda,\orcidlink{0000-0002-5861-9370}}
\myaffiliation{Institute of Space Science---INFLPR Subsidiary, Bucharest, Romania}

\author{Markus Prim\,\orcidlink{0000-0002-1407-7450}}
\myaffiliation{Universit\"at Bonn, Regina-Pacis-Weg 3, D-53113 Bonn, Germany}

\author{Michaela Queitsch-Maitland\,\orcidlink{0000-0003-4643-515X}}
\myaffiliation{University of Manchester, School of Physics and Astronomy, Schuster Building, Oxford Rd, Manchester M13 9PL, United Kingdom}

\author{Juan Rojo\,\orcidlink{0000-0003-4279-2192}}
\myaffiliation{Nikhef National Institute for Subatomic Physics, Science Park 105, 1098 XG Amsterdam, Netherlands}

\author{Hiroki Rokujo\,\orcidlink{0000-0002-3502-493X}}
\myaffiliation{Nagoya University, Furo-cho, Chikusa-ku, Nagoya 464-8602, Japan}

\author{Andr\'e Rubbia\,\orcidlink{0000-0002-5747-1001}}
\myaffiliation{Institute for Particle Physics, ETH Zurich, 8093 Zurich, Switzerland}

\author{Jorge Sabater-Iglesias\,\orcidlink{0000-0003-2328-1952}}
\myaffiliation{D\'epartement de Physique Nucl\'eaire et Corpusculaire, University of Geneva, CH-1211 Geneva 4, Switzerland}

\author{Osamu Sato\,\orcidlink{0000-0002-6307-7019}}
\myaffiliation{Nagoya University, Furo-cho, Chikusa-ku, Nagoya 464-8602, Japan}

\author{Paola Scampoli\,\orcidlink{0000-0001-7500-2535}}
\myaffiliation{Albert Einstein Center for Fundamental Physics, Laboratory for High Energy Physics, University of Bern, Sidlerstrasse 5, CH-3012 Bern, Switzerland}
\myaffiliation{Dipartimento di Fisica ``Ettore Pancini'', Universit\`a di Napoli Federico II, Complesso Universitario di Monte S.~Angelo, I-80126 Napoli, Italy}

\author{Kristof Schmieden\,\orcidlink{0000-0003-1978-4928}}
\myaffiliation{Universit\"at Bonn, Regina-Pacis-Weg 3, D-53113 Bonn, Germany}

\author{Matthias Schott\,\orcidlink{0000-0002-4235-7265}}
\myaffiliation{Universit\"at Bonn, Regina-Pacis-Weg 3, D-53113 Bonn, Germany}

\author{Anna Sfyrla\,\orcidlink{0000-0002-3003-9905}}
\myaffiliation{D\'epartement de Physique Nucl\'eaire et Corpusculaire, University of Geneva, CH-1211 Geneva 4, Switzerland}

\author{Davide Sgalaberna\,\orcidlink{0000-0001-6205-5013}}
\myaffiliation{Institute for Particle Physics, ETH Zurich, 8093 Zurich, Switzerland}

\author{Mansoora Shamim\,\orcidlink{0009-0002-3986-399X}}
\myaffiliation{CERN, CH-1211 Geneva 23, Switzerland}

\author{Savannah Shively\,\orcidlink{0000-0002-4691-3767}}
\myaffiliation{Department of Physics and Astronomy, University of California, Irvine, CA 92697-4575, USA}

\author{Yosuke Takubo\,\orcidlink{0000-0002-3143-8510}}
\myaffiliation{National Institute of Technology (KOSEN), Niihama College, 7-1, Yakumo-cho Niihama, 792-0805 Ehime, Japan}

\author{Noshin Tarannum\,\orcidlink{0000-0002-3246-2686}}
\myaffiliation{D\'epartement de Physique Nucl\'eaire et Corpusculaire, University of Geneva, CH-1211 Geneva 4, Switzerland}

\author{Ondrej Theiner\,\orcidlink{0000-0002-6558-7311}}
\myaffiliation{D\'epartement de Physique Nucl\'eaire et Corpusculaire, University of Geneva, CH-1211 Geneva 4, Switzerland}

\author{Simon Thor\,\orcidlink{0000-0002-9183-526X}}
\myaffiliation{Institute for Particle Physics, ETH Zurich, 8093 Zurich, Switzerland}

\author{Eric Torrence\,\orcidlink{0000-0003-2911-8910}}
\myaffiliation{University of Oregon, Eugene, OR 97403, USA}

\author{Oscar Ivan Valdes Martinez\,\orcidlink{0000-0002-7314-7922}}
\myaffiliation{University of Manchester, School of Physics and Astronomy, Schuster Building, Oxford Rd, Manchester M13 9PL, United Kingdom}

\author{Svetlana Vasina\,\orcidlink{0000-0003-2775-5721}}
\myaffiliation{Affiliated with an international laboratory covered by a cooperation agreement with CERN.}

\author{Benedikt Vormwald\,\orcidlink{0000-0003-2607-7287}}
\myaffiliation{CERN, CH-1211 Geneva 23, Switzerland}

\author{Yuxiao Wang\,\orcidlink{0009-0004-1228-9849}}
\myaffiliation{Department of Physics, Tsinghua University, Beijing, China}

\author{Eli Welch\,\orcidlink{0000-0001-6336-2912}}
\myaffiliation{Department of Physics and Astronomy, University of California, Irvine, CA 92697-4575, USA}

\author{Monika Wielers\,\orcidlink{0000-0001-9232-4827}}
\myaffiliation{Particle Physics Department, STFC Rutherford Appleton Laboratory, Harwell Campus, 
Didcot, OX11 0QX, United Kingdom}

\author{Benjamin James Wilson\,\orcidlink{0000-0002-7811-7474}}
\myaffiliation{University of Manchester, School of Physics and Astronomy, Schuster Building, Oxford Rd, Manchester M13 9PL, United Kingdom}

\author{Jialin Wu\,\orcidlink{0009-0001-6447-0410}}
\myaffiliation{Institute for Particle Physics, ETH Zurich, 8093 Zurich, Switzerland}

\author{Johannes Martin Wuthrich\,\orcidlink{0000-0002-7522-8160}}
\myaffiliation{Institute for Particle Physics, ETH Zurich, 8093 Zurich, Switzerland}

\author{Yue Xu\,\orcidlink{0000-0001-9563-4804}}
\myaffiliation{Department of Physics, University of Washington, PO Box 351560, Seattle, WA 98195-1460, USA}

\author{Samuel Zahorec\,\orcidlink{0009-0000-9729-0611}}
\myaffiliation{CERN, CH-1211 Geneva 23, Switzerland}
\myaffiliation{Charles University, Faculty of Mathematics and Physics, Prague, Czech Republic}

\author{Stefano Zambito\,\orcidlink{0000-0002-4499-2545}}
\myaffiliation{D\'epartement de Physique Nucl\'eaire et Corpusculaire, University of Geneva, CH-1211 Geneva 4, Switzerland}

\author{Shunliang Zhang\,\orcidlink{0009-0001-1971-8878}}
\myaffiliation{Department of Physics, Tsinghua University, Beijing, China}

\author{Xingyu Zhao\,\orcidlink{0000-0003-1348-5732} \PRE{\vspace*{0.1in}}}
\myaffiliation{Institute for Particle Physics, ETH Zurich, 8093 Zurich, Switzerland}

\begin{abstract}
The FASER experiment at CERN has opened a new window in collider neutrino physics by detecting TeV-energy neutrinos produced in the forward direction at the LHC. Building on this success, this document outlines the scientific case and design considerations for an upgraded FASER neutrino detector to operate during LHC Run 4 and beyond. The proposed detector will significantly enhance the neutrino physics program by increasing event statistics, improving flavor identification, and enabling precision measurements of neutrino interactions at the highest man-made energies. Key objectives include measuring neutrino cross sections, probing proton structure and forward QCD dynamics, testing lepton flavor universality, and searching for beyond-the-Standard Model physics. Several detector configurations are under study, including high-granularity scintillator-based tracking calorimeters, high-precision silicon tracking layers, and advanced emulsion-based detectors for exclusive event reconstruction. These upgrades will maximize the physics potential of the HL-LHC, contribute to astroparticle physics and QCD studies, and serve as a stepping stone toward future neutrino programs at the Forward Physics Facility.
\end{abstract}

\pagenumbering{gobble}
\maketitle

\begin{center}
\copyright~2025 CERN for the benefit of the FASER Collaboration. Reproduction of this article or parts of it is allowed as specified in the CC-BY-4.0 license.  
\end{center}

\clearpage
\section{Introduction and Scientific Context \label{sec:intro}}
\setcounter{page}{1} 
\pagenumbering{arabic}

\noindent \textbf{Physics in the Forward Region of the LHC:} Particle colliders are our primary tools to search for new physics at the energy frontier. The most energetic realization of this idea is the Large Hadron Collider (LHC), which collides protons with a center of mass energy of almost 14~TeV. Its large multi-purpose experiments were built around the LHC's interaction points (IPs) to analyze collision products with high transverse momentum in the central region, as, for example, expected from the decays of the Higgs boson or proposed new heavy particles at the electroweak scale. In addition, the LHC is also the source of the most energetic neutrinos produced by humankind. These are dominantly produced via the decay of hadrons and form an intense and tightly collimated beam in the forward direction, along the beam collision axis. Similarly, forward hadrons may also decay to as-yet-undiscovered light and weakly interacting particles, which are predicted by various models of new physics and could play the role of dark matter or be a mediator to the dark sector. Until recently, these particles escaped down the beam pipe and thereby evaded the LHC experiments. \medskip

\noindent \textbf{FASER in LHC Run~3:} The FASER experiment was designed and built to cover this \textit{blind spot} and exploit the opportunity for the first time. The experiment is situated about 480~m east of the ATLAS IP, inside a previously unused service tunnel, where it can be centered on the beam collision axis~\cite{FASER:2022hcn}. It currently consists of two sub-detectors. Placed on the upstream end is FASER$\nu$, a dedicated emulsion-tungsten neutrino detector with a target mass of about 1~ton. Located behind is the FASER magnetic spectrometer, which was primarily designed to search for decays of new particles but also acts as a spectrometer for the neutrino detector.  In March 2023, FASER detected the first 153 interactions of collider neutrinos through a muon appearance signature in the electronic detector, corresponding to a statistical significance of 16$\sigma$~\cite{FASER:2023zcr}. This result, along with 8 neutrinos subsequently detected by SND@LHC~\cite{SNDLHC:2023pun}, has established that TeV collider neutrinos can be seen with relatively fast, small, and inexpensive detectors. In March 2024, an analysis of the FASER$\nu$ data reported the first detection of electron neutrinos and the first measurement of $\nu_e$ and $\nu_\mu$ cross sections at TeV energies~\cite{FASER:2024hoe}. The first measurement of the muon neutrino flux using 338 events in the electronic detector was released in December 2024~\cite{FASER:2024ref}. \medskip

\noindent \textbf{Prospects for LHC Run~4:} The first observation of neutrinos at the LHC marks the \textit{dawn of collider neutrino physics}~\cite{Worcester:2023njy}. Although only a few hundred neutrinos have been observed and analyzed so far, FASER is projected to record around ten thousand neutrinos by the end of LHC Run~3 in mid-2026. Assuming a detector of similar size, several tens of thousands of neutrino interactions are expected during LHC Run~4 (2031 - 2034) and this figure increases to about a hundred thousand neutrinos at the end of the HL-LHC era (2031 - early 2040s). This large event rate will offer unique opportunities to study neutrinos at TeV energies, constrain proton structure, provide crucial input for astroparticle physics, and perform a variety of searches for new particles and phenomena. A summary of the physics potential is provided in \cref{sec:physics}. While FASER has been approved to operate during LHC Run~4~\cite{Boyd:2882503}, the collaboration plans to optimize the neutrino detector based on the experience learned during Run~3, as will be outlined in Section \ref{sec:detect} below. 

\section{Physics Objectives \label{sec:physics}}

The LHC produces an intense and strongly collimated beam of neutrinos and anti-neutrinos of all three flavors with TeV energies in the forward direction, some of which will interact in the FASER neutrino detector. This section will summarize the physics opportunities offered by measurements of these neutrinos. Several options for an upgraded neutrino detector to exploit this beam are considered in this document. Although their technologies, masses, and detector locations differ, their general physics objectives are qualitatively the same. When presenting quantitative results, we will consider a reference design for an on-axis electronic detector, as discussed in \cref{sec:on-axis}.

\begin{figure}[tbh]
\centering
\vspace*{-0.2cm}
\includegraphics[width=0.325\textwidth]{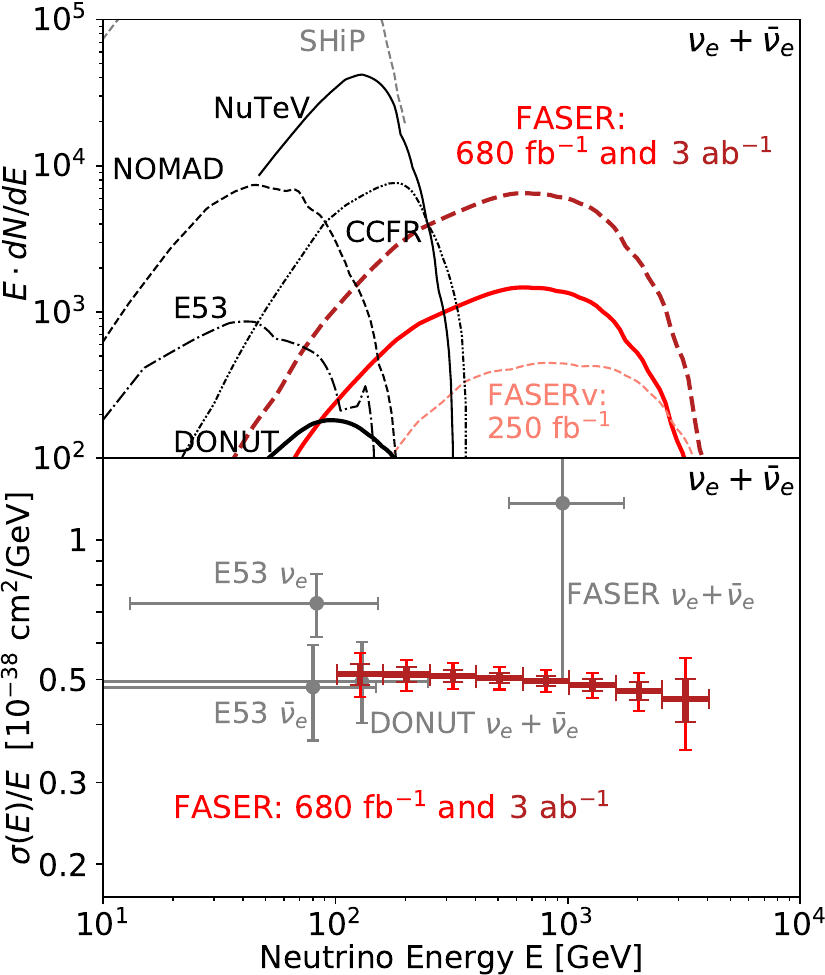}
\includegraphics[width=0.325\textwidth]{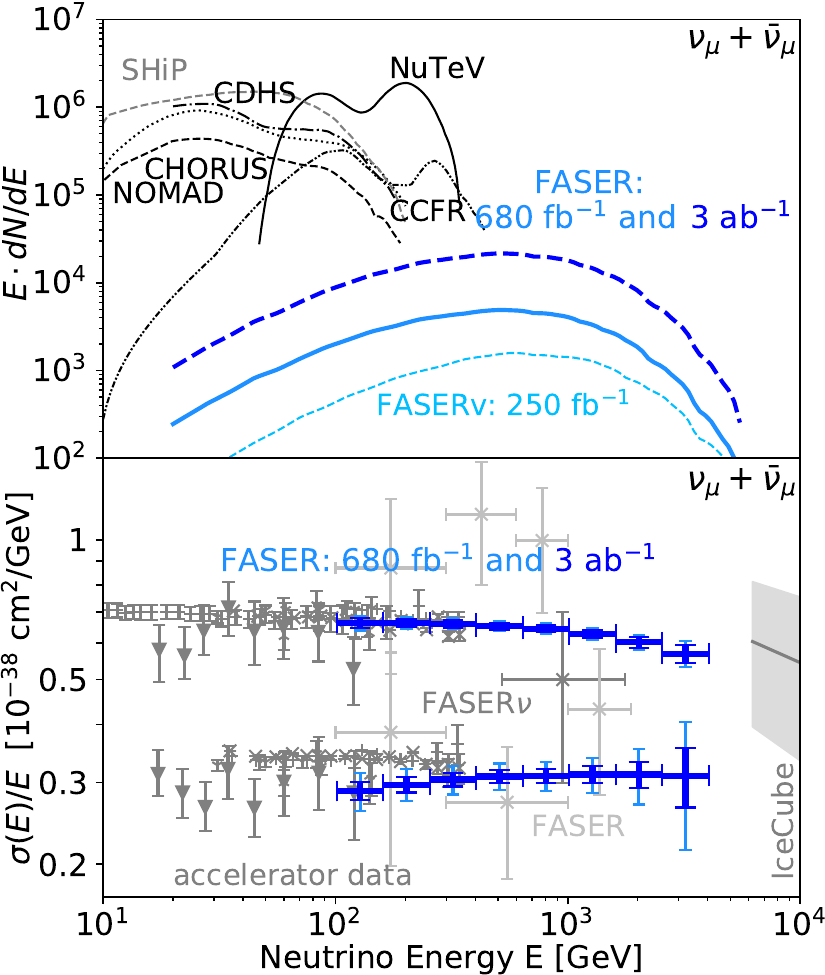}
\includegraphics[width=0.325\textwidth]{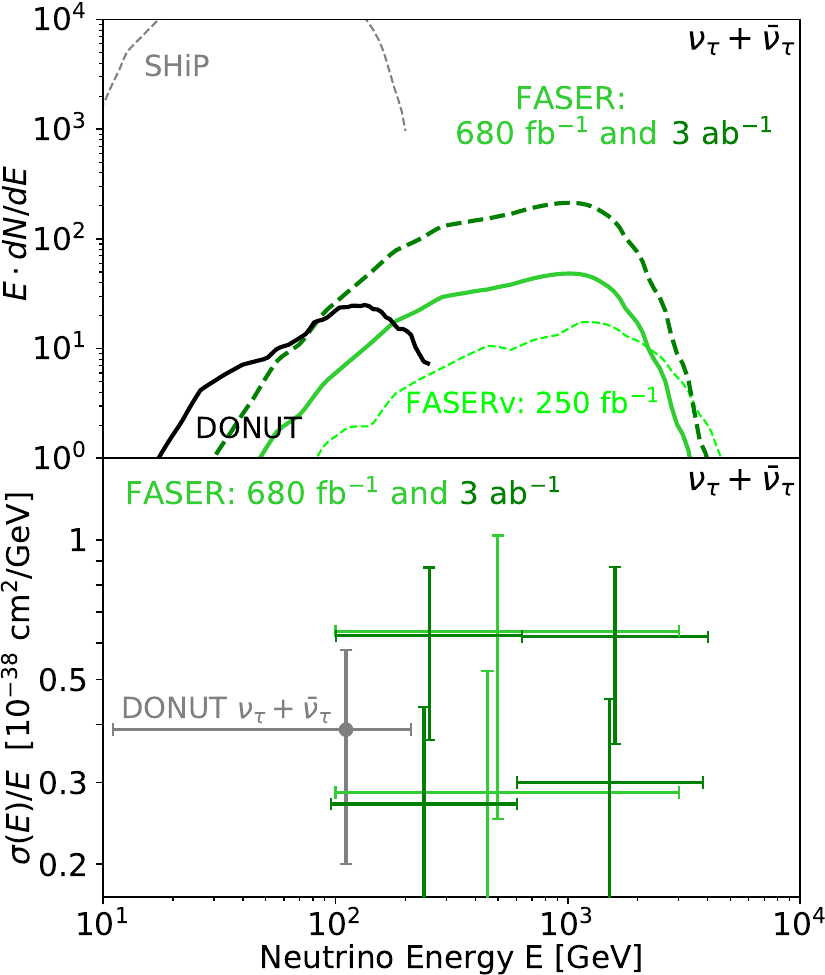}
\caption{The spectrum of neutrinos interacting in FASER (top) as a function of energy and the expected precision of FASER measurements of neutrino interaction cross sections (bottom, statistical errors only) for electron (left), muon (middle), and tau (right) neutrinos at FASER$\nu$ in LHC Run~3 ($250~\text{fb}^{-1}$) as well as an upgraded neutrino detector in LHC Run~4 ($680~\text{fb}^{-1}$) and the HL-LHC ($3~\text{ab}^{-1}$). The neutrino fluxes are estimated using EPOS-LHC and POWHEG+Pythia~\cite{FASER:2024ykc, Pierog:2013ria, Buonocore:2023kna}, and the cross section measurements assume a signal efficiency of 50\%. Existing data from accelerator experiments~\cite{ParticleDataGroup:2020ssz}, IceCube~\cite{IceCube:2017roe}, and the recent FASER results~\cite{FASER:2024hoe, FASER:2024ref} are shown, together with the projected fluxes for SHiP~\cite{Ahdida:2023okr}.}
\vspace*{-0.2cm}
\label{fig:fluxes}
\end{figure}

\noindent \textbf{Neutrino Event Rates:}   
In the following, we will assume a benchmark detector with dimensions of $25~\cm \times 60~\cm \times 1~\m$, chosen to fit the existing trench, consisting of a dense tungsten target interleaved with active detector elements and providing a total target mass of 1~ton. Using the state-of-the-art neutrino fluxes estimated with EPOS-LHC~\cite{Pierog:2013ria} and POWHEG+Pythia~\cite{Buonocore:2023kna} as discussed in Ref.~\cite{FASER:2024ykc}, about 3.2k electron neutrinos, 13k muon neutrinos, and 100 tau neutrinos are expected to interact inside the detector during LHC~Run~4 (with an expected luminosity of $680~\ifb$). This number will increase to about 14k electron neutrinos, 60k muon neutrinos, and 500 tau neutrinos when the detector is operated during the entire HL-LHC era (with a luminosity of $3~\iab$). The corresponding energy spectrum is shown in the upper panels of \cref{fig:fluxes}. One can see that FASER will probe the energy region from hundreds of GeV to several TeV, between accelerator experiments and astroparticle observatories like IceCube, and with an event rate that significantly exceeds those of FASER$\nu$ in Run~3 (with an assumed luminosity $250~\ifb$). \smallskip

\noindent \textbf{Neutrino Physics at TeV Energies:}  The large statistics imply the potential for precision studies. For example, assuming a well-known flux, FASER will be able to measure neutrino interaction cross sections at TeV energies with unmatched precision. This is illustrated in the lower panels of \cref{fig:fluxes}. Since FASER will measure the neutrino cross sections for all flavors, a comparison of the results will also allow probing lepton flavor universality in the neutrino scattering and complementing measurements performed by flavor physics experiments and the main LHC detectors. Alternatively, assuming the cross section predicted by the Standard Model (SM), FASER measurements will constrain the neutrino flux and constrain the modeling of forward particle production and proton structure. Finally, if one assumes both a well-known flux and interaction cross section, one can probe for BSM effects in neutrino production, such as decays of new particles into neutrinos~\cite{Kling:2020iar}, propagation, such as sterile neutrino-induced oscillations~\cite{FASER:2019dxq}, or interactions, such as non-standard neutrino interactions (NSI)~\cite{Falkowski:2021bkq}. 

Of course, event rates constrain a combination of neutrino fluxes, propagation, and interaction cross sections.  With the many events available, the full dataset will allow one to form differential distributions (for example, with respect to neutrino energy, pseudorapidity ($\eta$) or lepton momentum) or constrain exclusive channels (such as charm-associated neutrino interactions), which can be used to disentangle flux and interaction effects. For instance, the lepton momentum can reconstruct kinematic variables of charged-current (CC) deep-inelastic scattering (DIS) events, which are sensitive to partonic structure, exclusive channels like $\nu_\mu s \to \mu c$ probe specific initial quark flavors, while others with well-known cross sections can act as standard candles for flux measurements~\cite{Wilkinson:2023vvu}. \smallskip

\begin{figure}[tbh]
\centering
\vspace*{-0.6cm}
\includegraphics[width=0.99\textwidth]{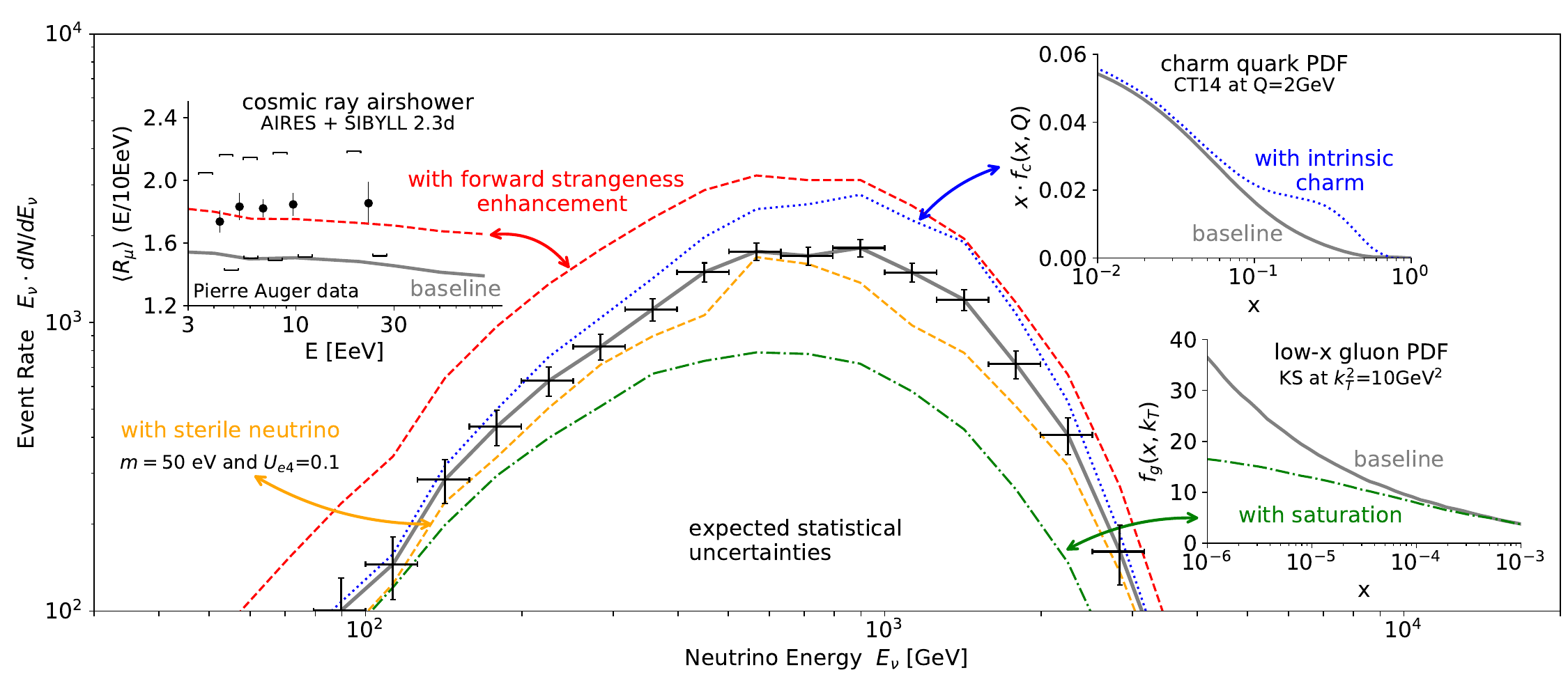}
\caption{The energy distribution of interacting electron neutrinos in FASER at LHC Run 4 with $680~\text{fb}^{-1}$. The baseline prediction, shown as gray solid line, uses SIBYLL~2.3d~\cite{Riehn:2019jet} to model light hadron production and $k_T$-factorization approach discussed in Ref.~\cite{Bhattacharya:2023zei} for charm production.  The colored lines illustrate four examples of physics that modify the spectrum:~(1) {\it neutrino oscillations} due to a sterile neutrino and the parameters indicated (orange dashed); (2) an {\it intrinsic charm} component in the proton, following the BHPS model~\cite{Brodsky:1980pb} implemented in the CT14 PDF~\cite{Hou:2017khm} as estimated in Ref.~\cite{Maciula:2022lzk} (blue dotted); (3) {\it gluon saturation} at low $x$~\cite{Kutak:2012rf}, as estimated in Ref.~\cite{Bhattacharya:2023zei} (green dash-dotted), and (4) {\it strangeness enhancement} introduced in Ref.~\cite{Anchordoqui:2022fpn} to resolve the cosmic-ray muon puzzle (red dashed).  The variations from these effects are all far greater than the expected statistical uncertainty (black error bars, assuming a 50\% signal efficiency).} 
\vspace*{-0.2cm}
\label{fig:Spectrum}
\end{figure}

\noindent \textbf{QCD and Proton Structure:} FASER will also enable unique studies of the strong interaction and proton structure both via measurements of neutrino fluxes and interactions. The former uses the fact that collider neutrinos mainly originate from the decay of charged pions, kaons, and charm mesons. While forward production of these particles at LHC energies has not been measured before, neutrino flux measurements provide a novel way to constrain forward hadron production and its underlying physics. This is illustrated in \cref{fig:Spectrum}, which shows the expected energy spectrum of interacting electron neutrinos as well as several phenomena that can modify it. The highest energy electron neutrinos primarily originate from charm hadrons. These are mainly produced via gluon fusion with one gluon carrying a momentum fraction $x\sim 1$ and the other $x \sim 4 m_c / s \sim 10^{-7}$. Thus, neutrino flux measurements will be sensitive to proton structure phenomena at high-$x$, such as intrinsic charm~\cite{Maciula:2022lzk}, and low-$x$, such as BFKL dynamics and gluon saturation~\cite{Bhattacharya:2023zei}, as shown in \cref{fig:Spectrum}. More generally, these measurements will allow the gluon parton distribution functions (PDFs) in the otherwise inaccessible low-$x$ region to be constrained, as shown in the left panel of \cref{fig:PDF}. 

\begin{figure}[tbh]
\centering
\vspace*{-2mm}
\includegraphics[width=0.41\textwidth]{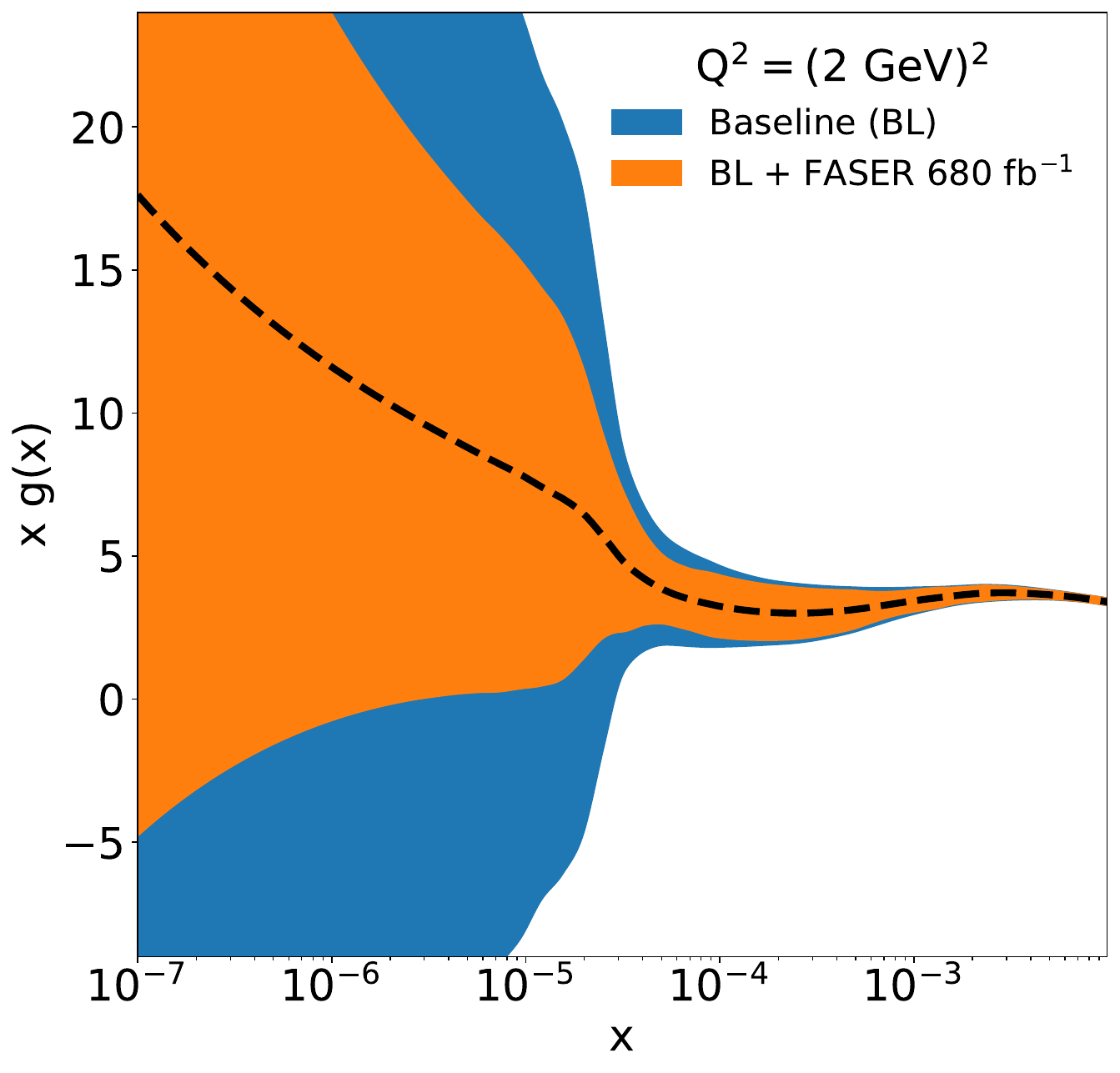}
\includegraphics[width=0.42\textwidth]{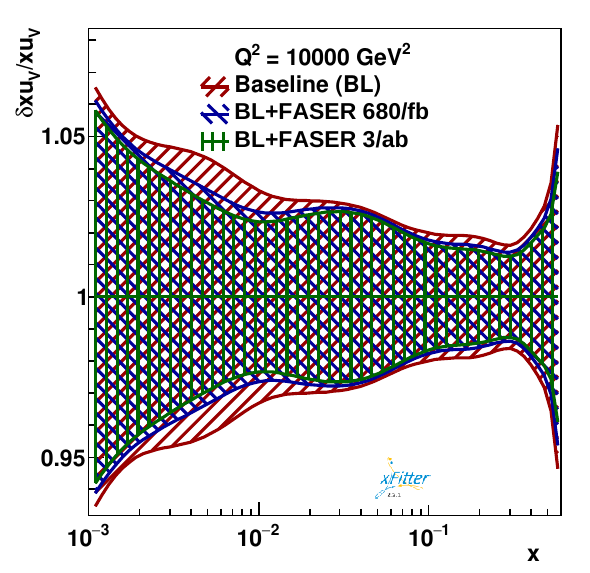}
\caption{Improvements due to FASER flux measurements on the small-$x$ gluon PDF (left) and interaction measurements on the valence down quarks (right). }
\vspace*{-2mm}
\label{fig:PDF}
\end{figure}

FASER also augments the LHC program with a \textit{neutrino-ion collider} to probe CC DIS events. This complements the planned electron-ion collider, which will probe neutral current (NC) DIS in a similar energy range. The large neutrino sample enables differential cross-section measurements to constrain PDFs. An example is shown in the right panel of \cref{fig:PDF} obtained considering statistical uncertainties only. These measurements will represent a non-trivial proof-of concept that LHC neutrino differential measurements can be unfolded to the cross-section level to be used in theoretical interpretations, paving the way and demonstrating the feasibility of subsequent neutrino DIS measurements at the FPF experiments. Such measurements will benefit key measurements at the LHC central detectors, such as Higgs or weak boson production~\cite{Cruz-Martinez:2023sdv}, and help to break the degeneracy between QCD and new physics effects in LHC data interpretations~\cite{Hammou:2024xuj}. \medskip \\

\noindent \textbf{Astroparticle Physics:} Improving forward particle production models is essential for astroparticle physics, where they are used to describe the production of high-energy particles in extreme astrophysical environments and the interactions of cosmic rays with Earth’s atmosphere. Neutrino flux measurements at FASER provide a unique opportunity to test these models under controlled experimental conditions. For example, FASER will help shed light on the muon puzzle, which refers to the fact that, for many years, experimental measurements of the number of muons in high- and ultra-high-energy cosmic-ray air showers have been in tension with model predictions~\cite{Soldin:2021wyv}. Extensive studies have suggested that an enhanced rate of strangeness production in the forward direction could explain the discrepancy~\cite{Albrecht:2021cxw}. If the muon puzzle is resolved by enhanced strangeness, which may lead to differences in the predicted neutrino fluxes that exceed a factor of two~\cite{Anchordoqui:2022fpn}, this will be evident in FASER measurements of the electron neutrino spectrum, as shown in \cref{fig:Spectrum}. In addition, neutrino flux measurements will also reduce uncertainties in the prompt atmospheric neutrino flux. This flux, arising from charmed hadron decays in cosmic ray collisions with the atmosphere, becomes the dominant background for astrophysical neutrino searches at energies above a few 100~TeV. Refined predictions guided by FASER data will enhance astrophysical neutrino studies and multi-messenger astronomy~\cite{Bai:2022xad}. \medskip

\noindent \textbf{New Particles:} The FASER neutrino detector may add considerable discovery potential to the FASER experiment reach for dark matter and other BSM particles in regions of parameter space that are currently unprobed. In addition to decays signatures of long-lived particles, such as dark photons~\cite{FASER:2023tle} and ALPs~\cite{FASER:2024bbl} that are the target of the main FASER detector~\cite{Boyd:2882503}, the neutrino detector offers the opportunity to search for scattering signatures. For example, milli-charged particles, which have a variety of motivations and may be a consequence of a dark sector with an unbroken dark electromagnetism, can scatter in FASER, producing recoiling electrons with energies in the 300~MeV to 10~GeV range. The flux of milli-charged particles is maximal in the far-forward region, and the $N_{obs}=3$ sensitivity contour extends well into currently viable regions of parameter space as shown in the left panel of \cref{fig:BSM}. Another example is a sterile neutrino $N$ which couples through a dipole portal interaction $\frac{1}{2}\mu_{\nu}\bar{\nu}_L\sigma^{\alpha\beta} N F_{\alpha\beta}$. This induces elastic electron scattering $\nu_L e \to N_R e$, providing a signature similar to that of millicharged particles. Such events can lead to the discovery of new physics at FASER, see the right panel of \cref{fig:BSM}. 

\begin{figure}[thb]
\centering
\includegraphics[width=0.45\textwidth]{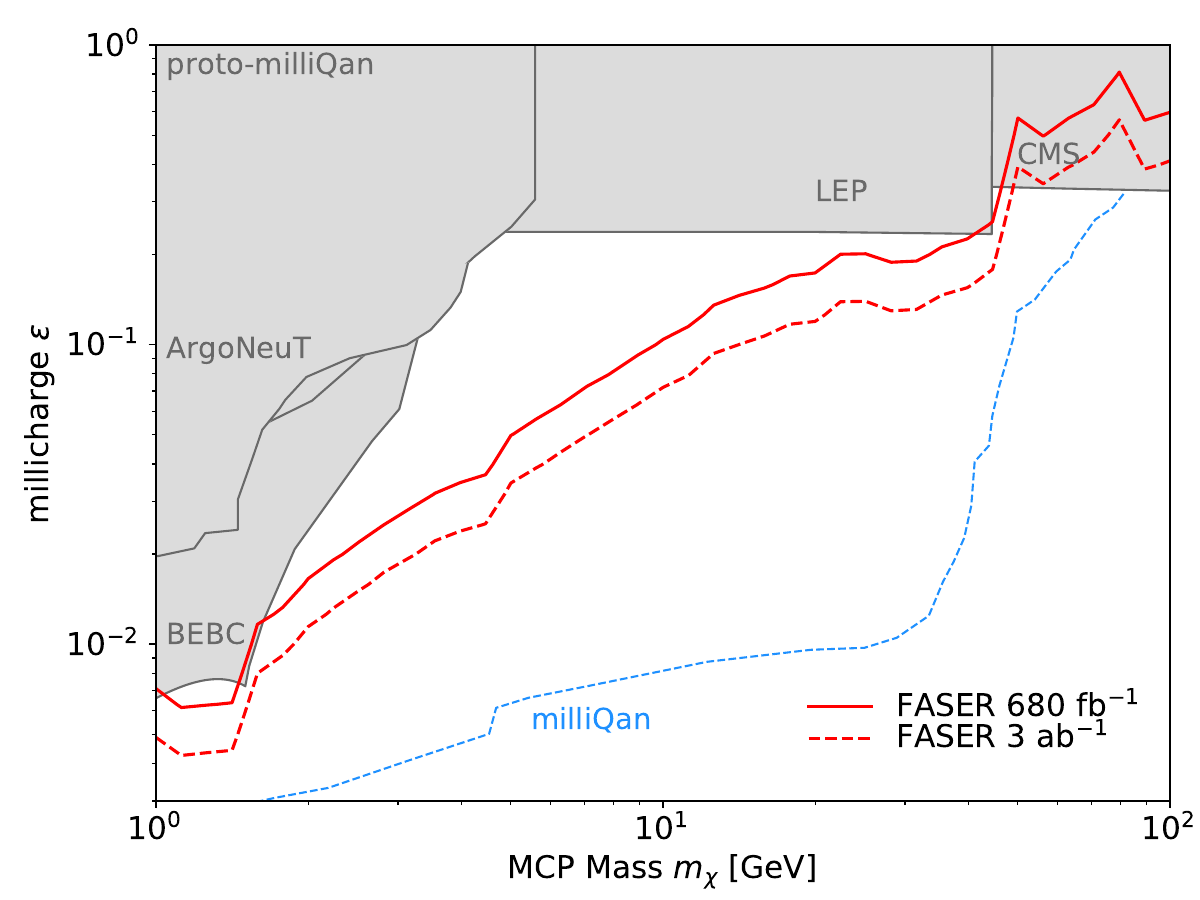}
\includegraphics[width=0.45\textwidth]{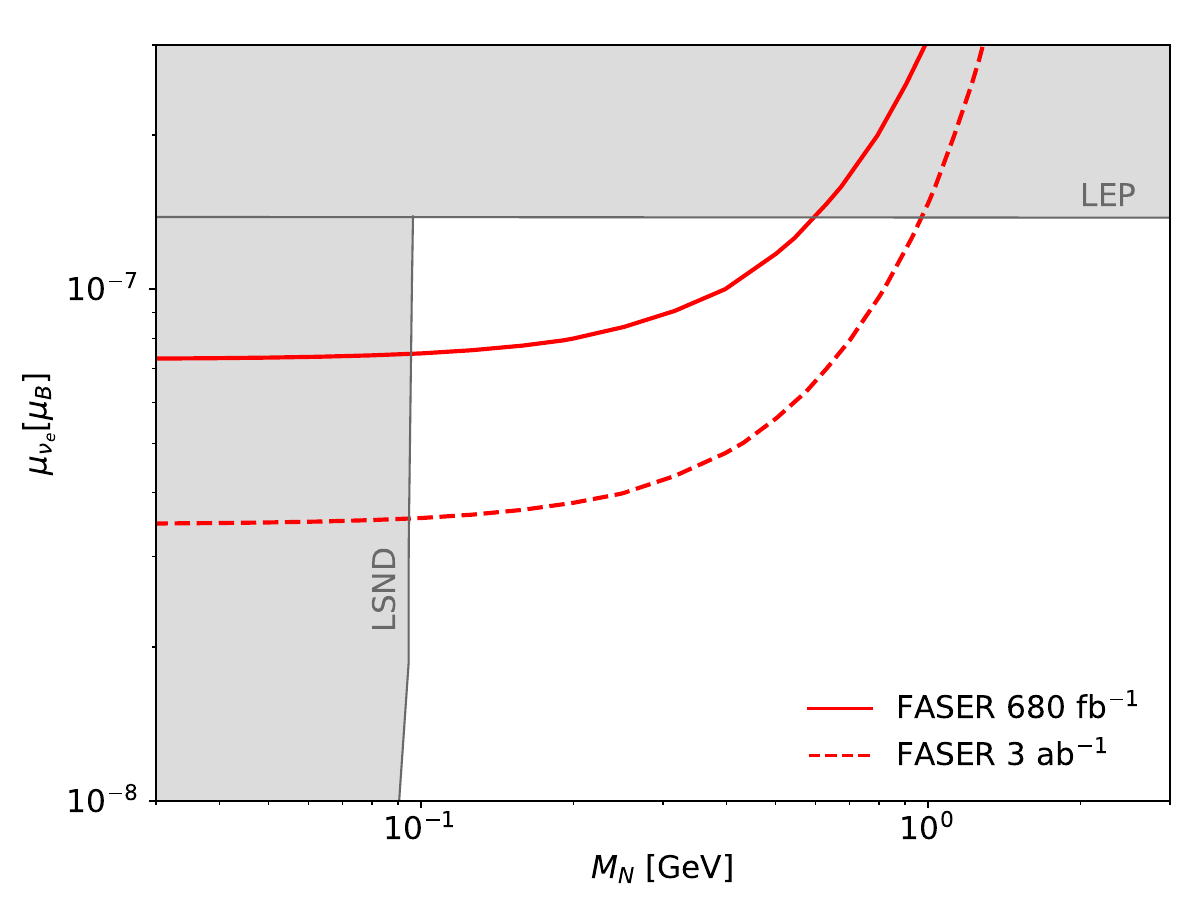}
\vspace*{-0.1in}
\caption{The discovery reach for milli-charged particles~\cite{Kling:2022ykt} (left) and heavy neutral leptons coupling to the SM through the magnetic dipole operator~\cite{Ismail:2021dyp} (right) for LHC Run 4 ($680~\text{fb}^{-1}$) and the HL-LHC ($3~\text{ab}^{-1}$). The contours are $N_{obs}=3$ signal contours, and the signal is single electron recoils with energies from 300 MeV to 10 GeV.}
\vspace*{-0.1in}
\label{fig:BSM}
\end{figure}

\section{Upgrade of the FASER Neutrino Detector \label{sec:detect}}

\noindent \textbf{General considerations:} 
The FASER experiment~\cite{Feng:2017uoz, FASER:2019dxq}, a cost-effective initiative proposed in 2018 and approved in 2019, is located 480 meters downstream of ATLAS and began data collection at the start of LHC Run~3 in 2022. FASER has been strategically placed to explore feebly interacting particles and neutrinos within a very low background environment~\cite{FASER:2018eoc}. Beam-related backgrounds, such as muons or neutral hadrons, can be mitigated with proper vetoing systems. FASER has functioned seamlessly during the 2022, 2023 and 2024 LHC data-taking periods, successfully collecting 97\% of the luminosity delivered to IP1, and it will continue to operate for the remainder of Run~3 (until mid 2026). FASER has received approval to continue its operations during the Run~4 period (2030-2033)~\cite{Boyd:2882503}.

The FASER detector was designed to meet tight timelines and operate within a limited budget, anticipating low particle rates and reduced radiation compared to the significantly larger LHC experiments. The setup includes 0.6~T permanent magnets, with the first magnet forming a 1.5-meter-long decay volume. It features four tracking stations that use ATLAS semiconductor tracker modules \cite{ATLAS:2007kpw}, scintillator stations that efficiently veto incoming muon particles and provide precise timing relative to the IP, a small calorimeter based on LHCb ECAL modules \cite{LHCb:2000vji}, and a passive emulsion-tungsten detector constituting the FASER$\nu$ component. This last component consists of 730 emulsion films interleaved with tungsten plates, resulting in a total active mass of about 1~ton.

The reference scenario for Run~4 is to consider electronic detectors, with the possibility of adding an emulsion target if compatible with the Run~4 environment (see Section \ref{sec:emulsionsrun4}). Neutrino interactions will occur within a possible target to be placed in front of FASER or in the rest of the FASER detector material, and $\nu_\mu$ scattering can remain detectable via FASER's electronic detectors, as demonstrated in Ref.~\cite{FASER:2023zcr, FASER:2024ref}. However, without a detector upgrade, physics results will be systematically limited, and it will not be possible to perform any other detailed studies that were enabled by the emulsion component of the experiment. \medskip  

\noindent \textbf{Choice of detector design:} 
The design of the upgraded detector largely impacts its ability to identify neutrino interactions and measure associated kinematics. The neutrino physics program at the HL-LHC and the associated experimental requirements are summarized in Table \ref{tab:physics_objectives}. 

\begin{table}[h]
\centering
\begin{tabular}{p{5.75cm}|p{2.5cm}|c c c c c | p{4.5cm} }
    \hline \hline
    \multicolumn{8}{c}{\textbf{Run~4 FASER neutrino physics programme}}\\
    \hline \hline
    Physics Objective & Measurement & $\nu_e$ & $\nu_\mu$ & $\nu_\tau$ & $q_\mu$ & $\eta_\nu$ & Additional Requirements \\
    \hline
    inclusive DIS cross section  & $\sigma_{CC}(E_\nu)$ & \checkmark & \checkmark & (\checkmark) & \checkmark &  & separately for all flavors \\
    \hline
    $\nu_\tau$ physics: test of lepton
    flavor universality and NSI & $\sigma_{CC}(E_\nu)$  & \checkmark & \checkmark & \checkmark & (\checkmark) &  & identify $\nu_\tau$ either statistically or event-by-event \\
    \hline
    NC scattering: cross section,
    NSI, $\theta_{W}$, charge radius... & $\sigma_{NC}(E_\nu)$ &  &  &  &  &  & identify NC events \\
    \hline
    constrain high-x PDFs & $d^2\sigma / dE_\mu d\theta_\mu$
    ($d^3\sigma / dx dy dQ$) &  & \checkmark &  &  \checkmark &  & ideally with charm ID 
    via neutrino scattering and large target mass \\
    \hline
    improve hadronic interaction 
    models, solve muon puzzle & $d^2\Phi/dE_\nu d\eta_\nu$ & \checkmark & \checkmark &  & \checkmark & \checkmark & neutrinos from pion/kaon decay ideally with extended $\eta$ coverage \\
    \hline
    constrain and test low-x QCD, prompt atmospheric neutrinos  & $d^2\Phi / dE_\nu d\eta_\nu$ & \checkmark &  & (\checkmark) & \checkmark & \checkmark & neutrinos from charm decay ideally with extended $\eta$ coverage\\
    \hline
    new particle searches & low recoil $e$ &  &  &  &  &  & event identification \\
    \hline\hline
\end{tabular}
\caption{Summary of physics objectives, required measurements, and experimental requirements. Here $\nu_\ell$, $q_\mu$ and $\eta_\nu$ mean that the measurements requires the ability to identify neutrinos of flavor $\ell = e, \mu, \tau$, to measure the muon charge, and to measure the neutrino rapidity, respectively.}
\label{tab:physics_objectives}
\end{table}

In order to achieve sufficient event rates for the physics goals, an upgraded neutrino detector should reach a target mass of about 1 ton, provide a sufficiently fine granularity to identify and separate the flavor of the incoming interacting neutrino with good $\nu_e$ vs. $\nu $ NC  separation (i.e. electron$/\pi^0$), and provide sufficient containment to contain electromagnetic and hadronic components of the event. Finally, the possibility of identifying muons is crucial. Detection of tau neutrinos, either statistically or directly via exclusive channels, remains a very challenging goal and is under investigation. \medskip  

\noindent \textbf{Detector location and space constraints:}
A map of the TI12 tunnel and UJ12 cavern is shown in \cref{fig:space}~(left).
The most efficient location in terms of rate per unit mass is along the current FASER experiment, which is aligned with the beam axis. The space in the trench in front of FASER becomes available if one removes the FASER$\nu$ detector (see central panel of \cref{fig:space}). The trench has been dug into the concrete floor and cannot be extended; therefore, any upgrade should fit within a maximal area of $50 \times 100$~cm$^2$.
\begin{figure}[htb]
\centering
\includegraphics[width=0.99\textwidth]{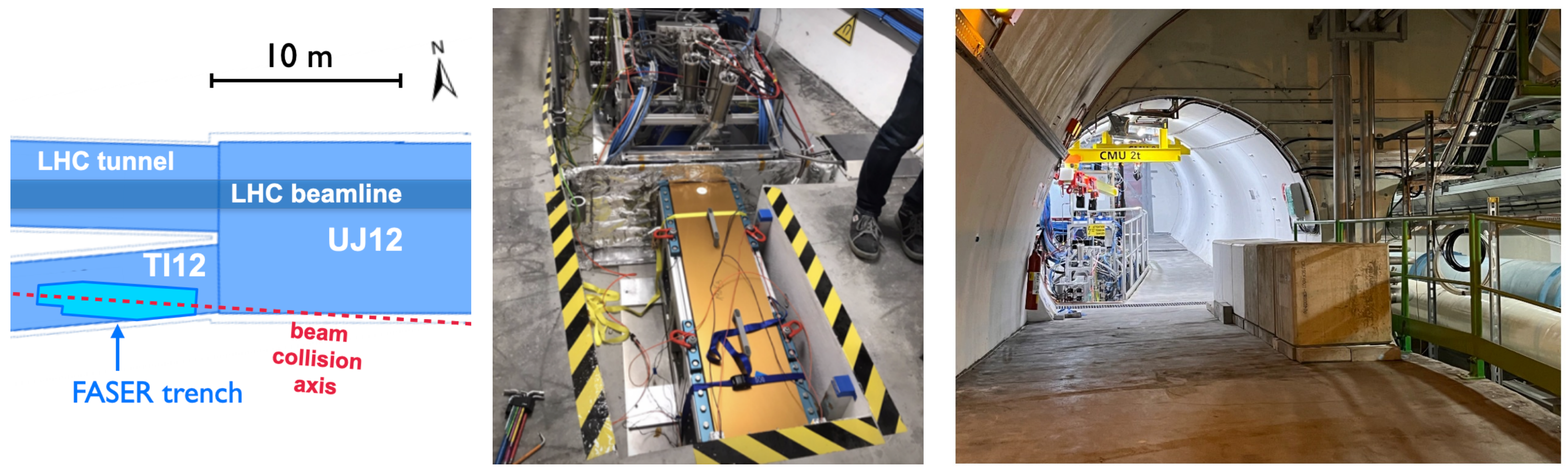}
\caption{
Map of the TI12 tunnel and UJ12 cavern (left); view of the trench with the FASER$\nu$ emulsion detector installed (center); and available space in the UJ12 tunnel behind FASER (right).} 
\label{fig:space}
\vspace*{-0.1in}
\end{figure}
The space behind the current FASER detector, within the UJ12 tunnel, will also be available to place an upgraded detector of potentially larger dimensions than the one in the trench (see right panel of \cref{fig:space}). 
The upgraded detector can be located slightly off-axis, resulting in lower neutrino flux, but providing the opportunity to study events at lower rapidity. The on-axis and off-axis locations provide complementary physics programs, providing a wide rapidity coverage, as well as the possibility to consider different detector optimizations to address in a complementary way the physics goals outlined in Table~\ref{tab:physics_objectives}.

\section{A reference design: electronic on-axis detectors}
\label{sec:on-axis}
\noindent In this section, we focus on options for on-axis detectors located in the trench.

\medskip
\noindent \textbf{The Forward TeV Neutrino detector (FORTVNE):}
The reference design for the upgraded detector, under further investigation, is a scintillator+tungsten tracking calorimeter. Its location and dimensions are shown in \figref{FORTVNE}. It measures approximately 1 m in depth ($Z$-direction), 25 cm in width ($X$-direction), and 60 cm in height ($Y$-direction). New scintillator bars will replace \fasernu's emulsion to allow electronic detection and event reconstruction. The detector consists of 66 modules along the beam direction, each containing a 5~mm-deep tungsten layer, a 5~mm-deep layer of horizontal scintillating bars for the $Y$-$Z$ view, and a 5~mm-deep layer of vertical scintillating bars for the $X$-$Z$ view. Each scintillator bar has two silicon photomultipliers (SIPMs) coupled to it, one at each end, to enable photon detection in both low-energy and high-energy ranges.  

\begin{figure}[htbp]
\centering
\includegraphics[width=0.44\textwidth]{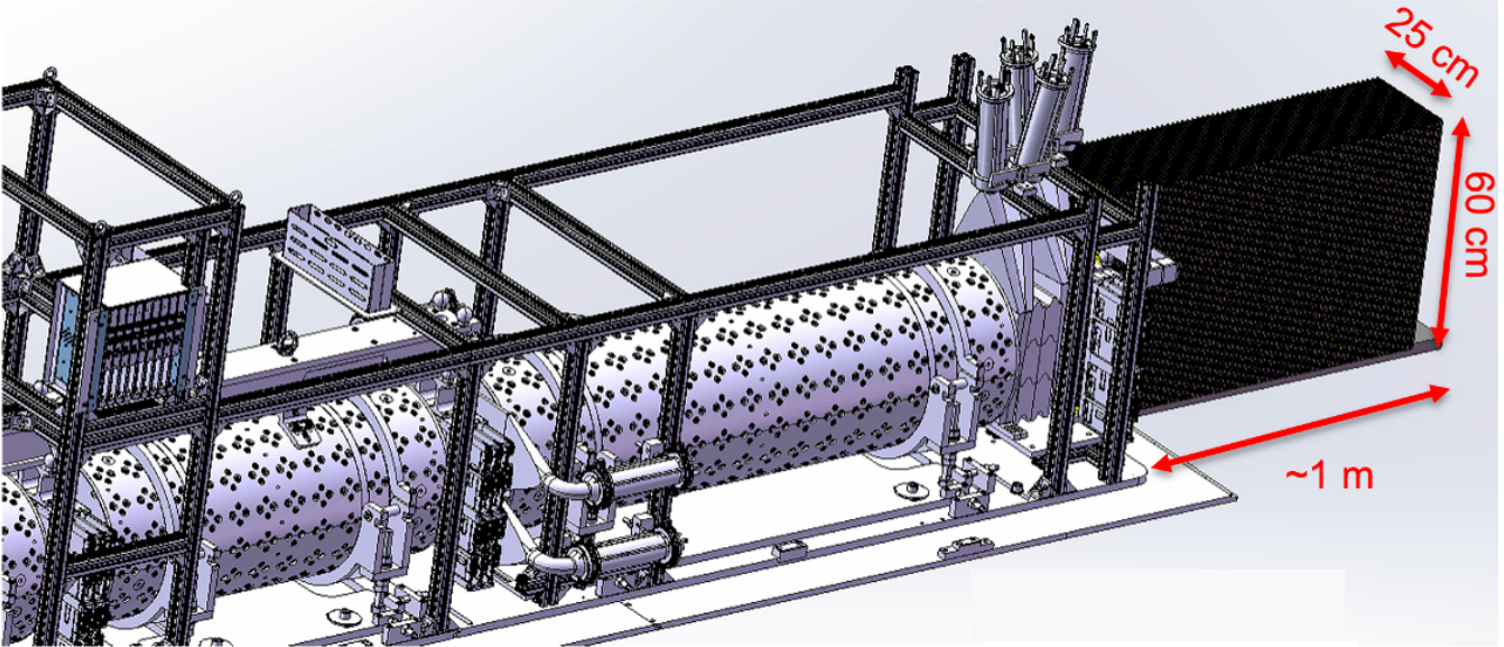}
\includegraphics[width=0.55\textwidth]{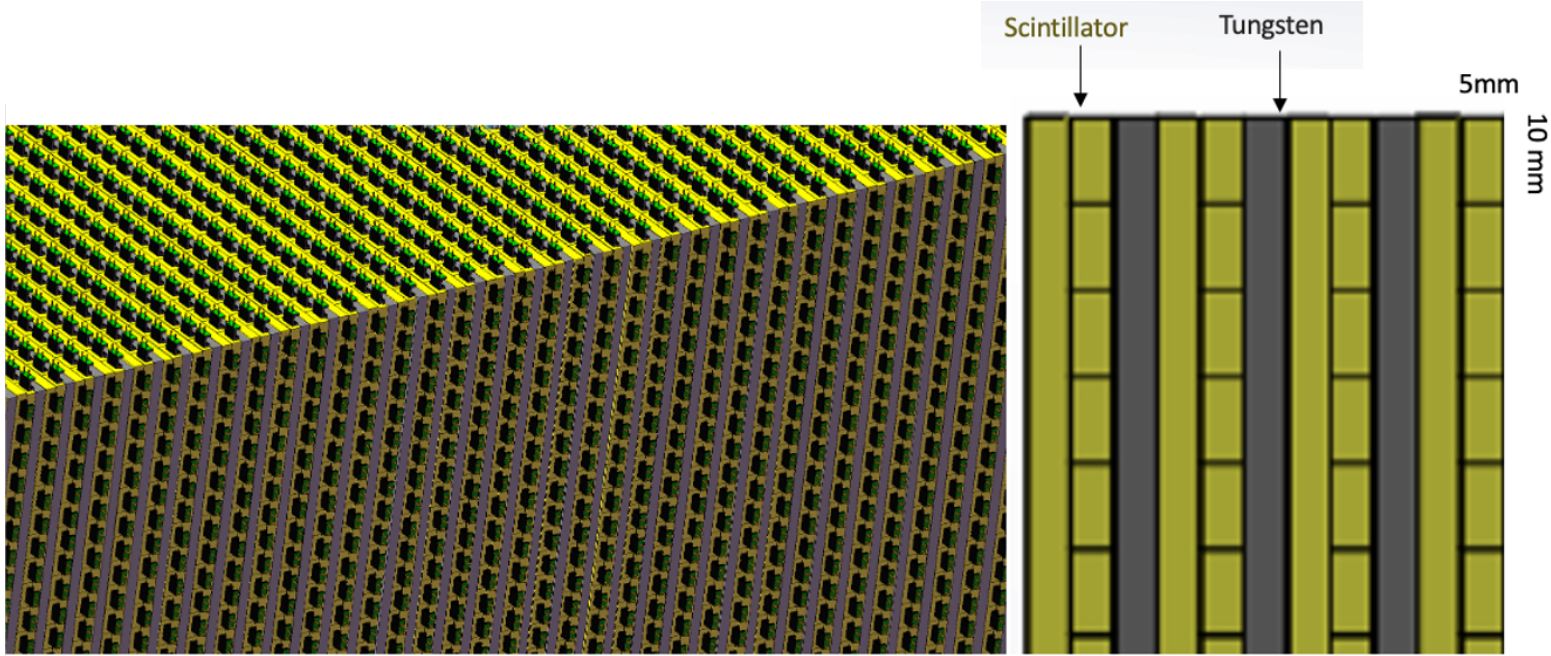}
\caption{(left) The FORTVNE detector (black) will be located upstream of the FASER spectrometer (gray) and has a total target mass of about 1 ton and dimensions of approximately $1~\text{m} \times 25~\text{cm} \times 60~\text{cm}$. (right) FORTVNE is 66 repeating modules, with each module consisting of 5 mm-thick tungsten plates, 5 mm-thick horizontal scintillator bars, and 5 mm-thick vertical scintillator bars.} 
\label{fig:FORTVNE}
\vspace*{-0.1in}
\end{figure}

The detector has a total of 1,650 vertical scintillator bar + SiPM assemblies and 3,960 horizontal scintillator bar + SiPM assemblies, resulting in 5,610 assemblies and 11,220 electronic readout channels overall. The total mass of the detector is approximately 1 ton. 

To increase the detector's mass and increase the detector's pseudorapidity ($\eta$) coverage, its height has been increased compared to \fasernu. The primary goal of the detector is to detect $\nu_e$ CC events, so achieving reasonable granularity in the scintillator bars and maintaining a tungsten thickness of less than 2 radiation lengths is essential to capture the development of electron showers. Some simulated events in the setup are shown in Figure \ref{fig:evd}. These clearly demonstrated the event and particle identification capabilities.\medskip  

\begin{figure}[htbp]
\centering
\includegraphics[width=0.98\textwidth]{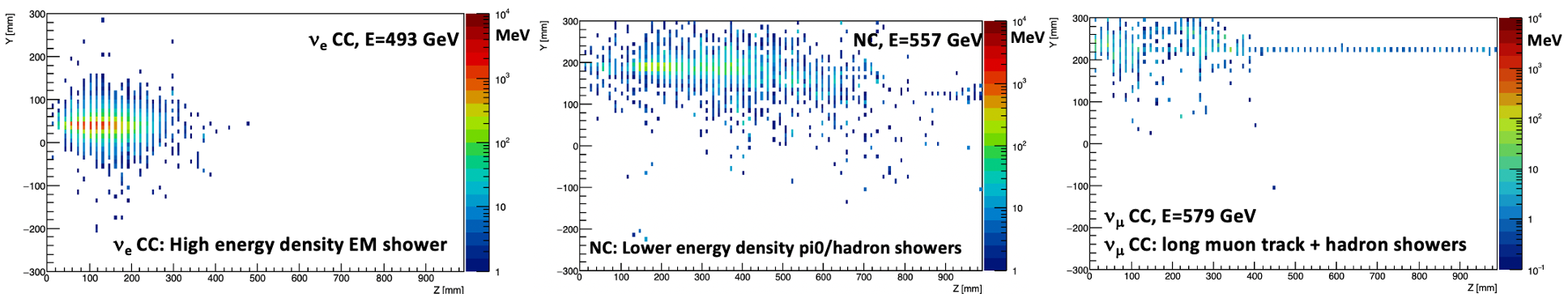}
\caption{Examples of simulated events in FORTVNE ($Y$-$Z$ view): (left) $\nu_e$ CC, (middle) NC, (right) $\nu_\mu$ CC.} 
\label{fig:evd}
\vspace*{-0.1in}
\end{figure}

\begin{wrapfigure}{r}{0.43\textwidth}
\centering

\includegraphics[width=0.38\textwidth]{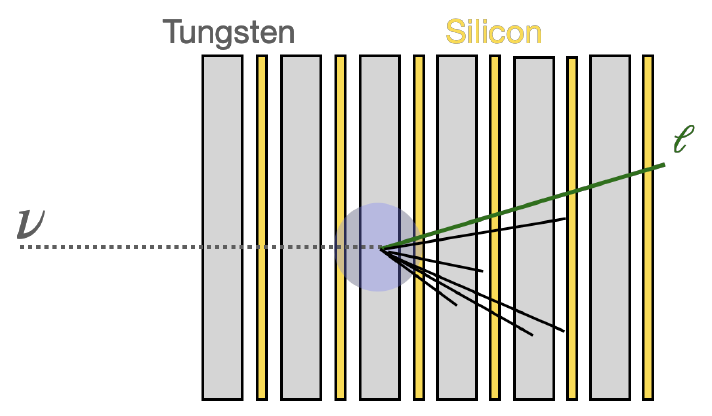}
\includegraphics[width=0.34\textwidth]{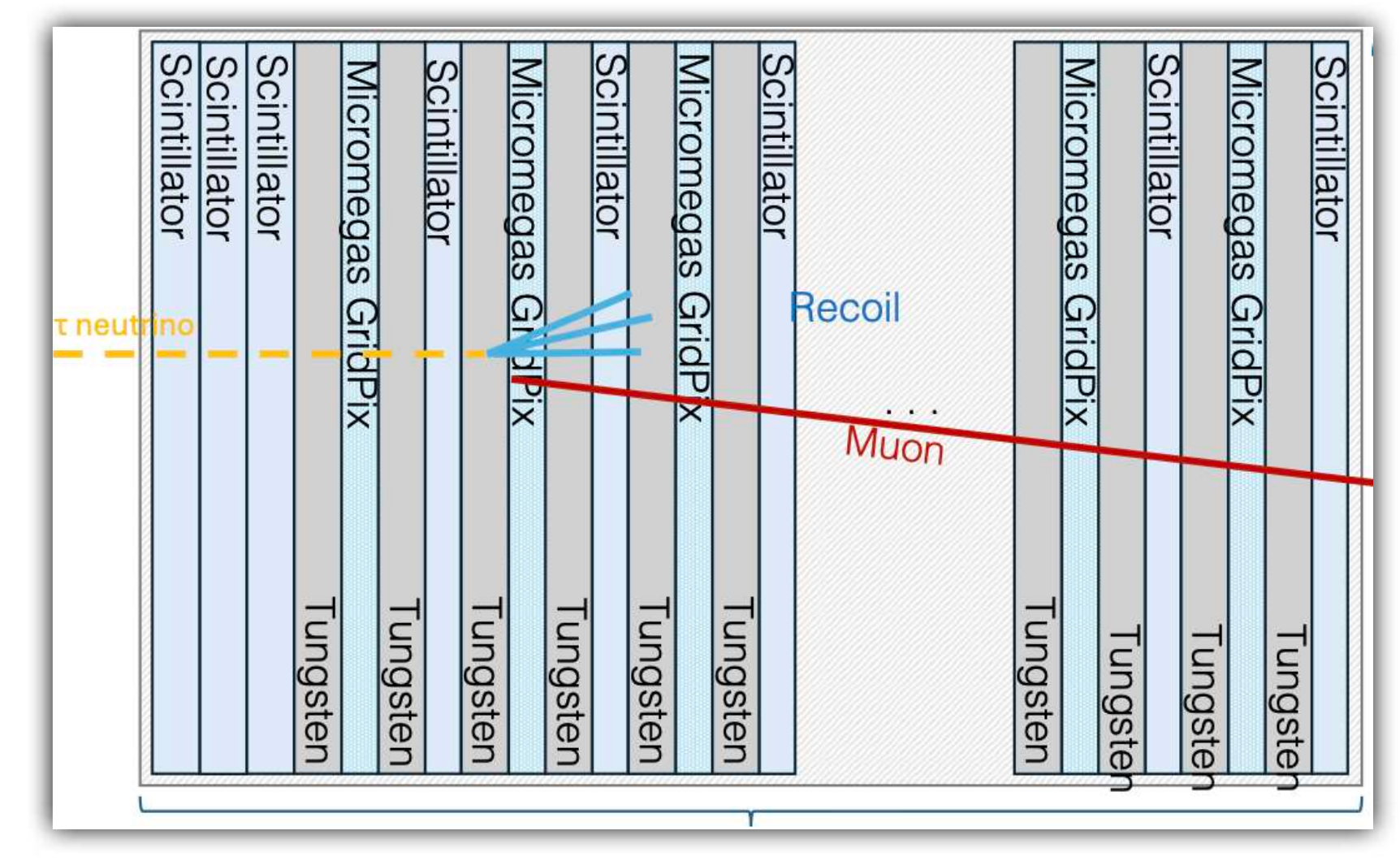}
\caption{Concept of a scintillator + $W$ detector complemented with high-precision tracker layers.
(left) With silicon strip detectors (e.g. ATLAS SCT); (right) With Micromegas Gridpix detector as high precision pixel detectors.} 
\label{fig:highprecisiontracker}
\vspace*{-0.1in}
\end{wrapfigure}

\noindent \textbf{Improvements with high precision trackers:}
Although the segmented scintillator bars provide tracking and calorimetry, their resolution is limited by their cm-scale granularity. 
The average kink angle of tau decays is expected to be only 2~mrad.
The scintillator granularity is not sufficient to reconstruct the secondary vertices and the impact parameters of tracks at the primary vertex, to efficiently identify the presence of the tau lepton or charmed hadrons. 

Options where the scintillator bars and $W$ target are complemented by layers of high-precision trackers, composed of silicon strips or pixel detectors, are shown in \cref{fig:highprecisiontracker}. 
 We are investigating the possibility of recycling the ATLAS SCT modules~\cite{ATLAS:2007kpw}. Using the modules from the two least irradiated layers would provide a total of 1,056 modules.
The neutrino detector would consist of alternating layers of tungsten and silicon, with eight SCT modules per layer.
An alternative option considers GridPix detectors~\cite{Kaminski:2017bgj} to provide a high precision pixel layer.
GridPix with TimePix-3 readout is based on the Micromegas principle, incorporating pixel readout with time information. It serves as the baseline technology for the babyIAXO experiment~\cite{IAXO:2020wwp}. The system offers high precision, with a per-layer position resolution of $20-30\, \mu$m and a tracklet angle resolution of less than $3.5$~mrad. Simulation studies are presently being performed to optimize and assess the performance of such layouts.

\section{Options for off-axis detectors\label{sec:xx}}

\noindent In this section, we discuss complementary detectors that could be located off-axis to further enhance the physics program.

\medskip

\noindent \textbf{High-granularity calorimeters for the CEPC:}
The Sci-W ECAL and the AHCAL are two existing prototypes built in the context of high-granularity calorimeters for the CEPC. Their designs, illustrated in Figure \ref{fig:ECALAHCAL}, are respectively $W$-scintillator pad and iron-scintillator pad detectors~\cite{Duan:2021mvk, Shi:2022fsb, Niu:2025xyu, 10646562, Xia:2025lhu}.
\clearpage

\begin{wrapfigure}{r}{0.5\textwidth}
\vspace{-2mm}
\includegraphics[width=0.5\textwidth]{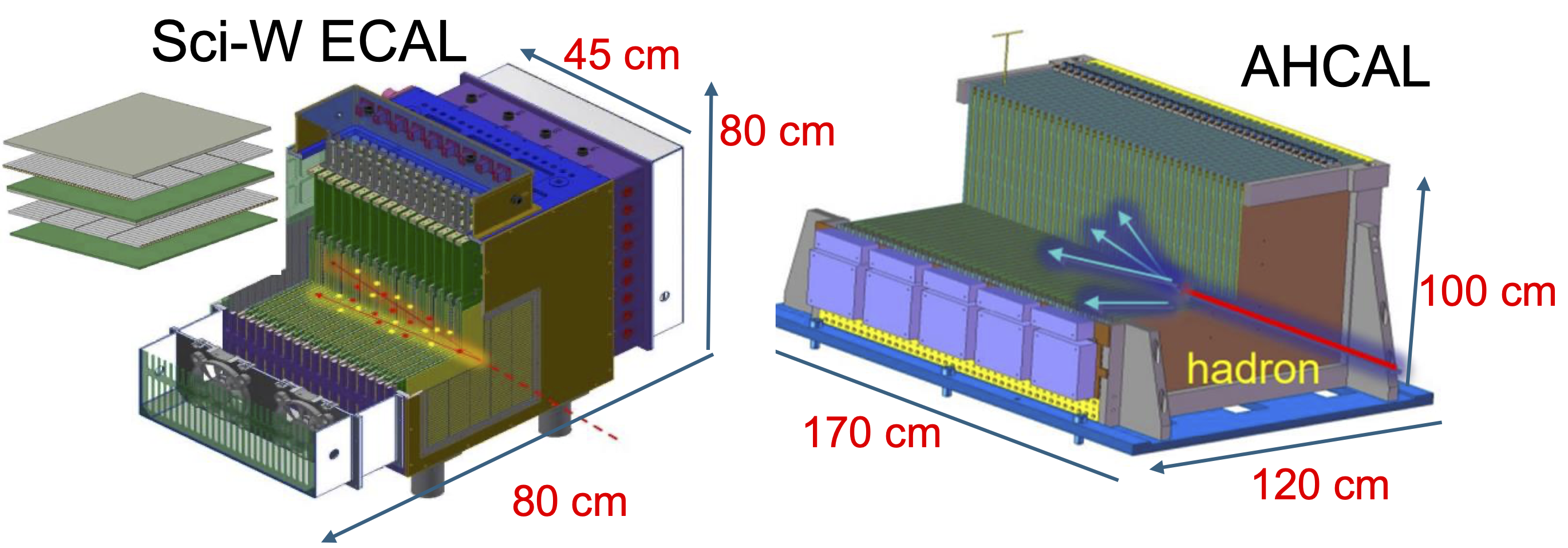}
\caption{Illustration of the Sci-W ECAL and AHCAL detectors. } \label{fig:ECALAHCAL}
\vspace{-6mm}
\end{wrapfigure}

They have been optimized for the particle flow Algorithm (PFA) and would provide calorimetric reconstruction of neutrino interactions with fine segmentation. These prototypes or a detector based on the same concept could be located behind the current FASER detector and detect neutrino interactions with sufficient resolution to discriminate between electron and muon neutrino CC events. 
\vspace{0.8cm} 

\noindent \textbf{3D optically isolated scintillator cubes, FASERCal:}
A custom design for an optimized detector system is considered, which integrates technologies proven with the SuperFGD (SFGD) detector at the T2K near detector \cite{T2K:2019bbb, Blondel:2017orl, Blondel:2020hml} and are currently being further developed within the CERN 3DET Collaboration, to obtain 3D scintillating voxels \cite{Boyarintsev:2021uyw}.

\begin{wrapfigure}{r}{0.6\textwidth}
\vspace{-3.5mm}
\includegraphics[width=0.6\textwidth]{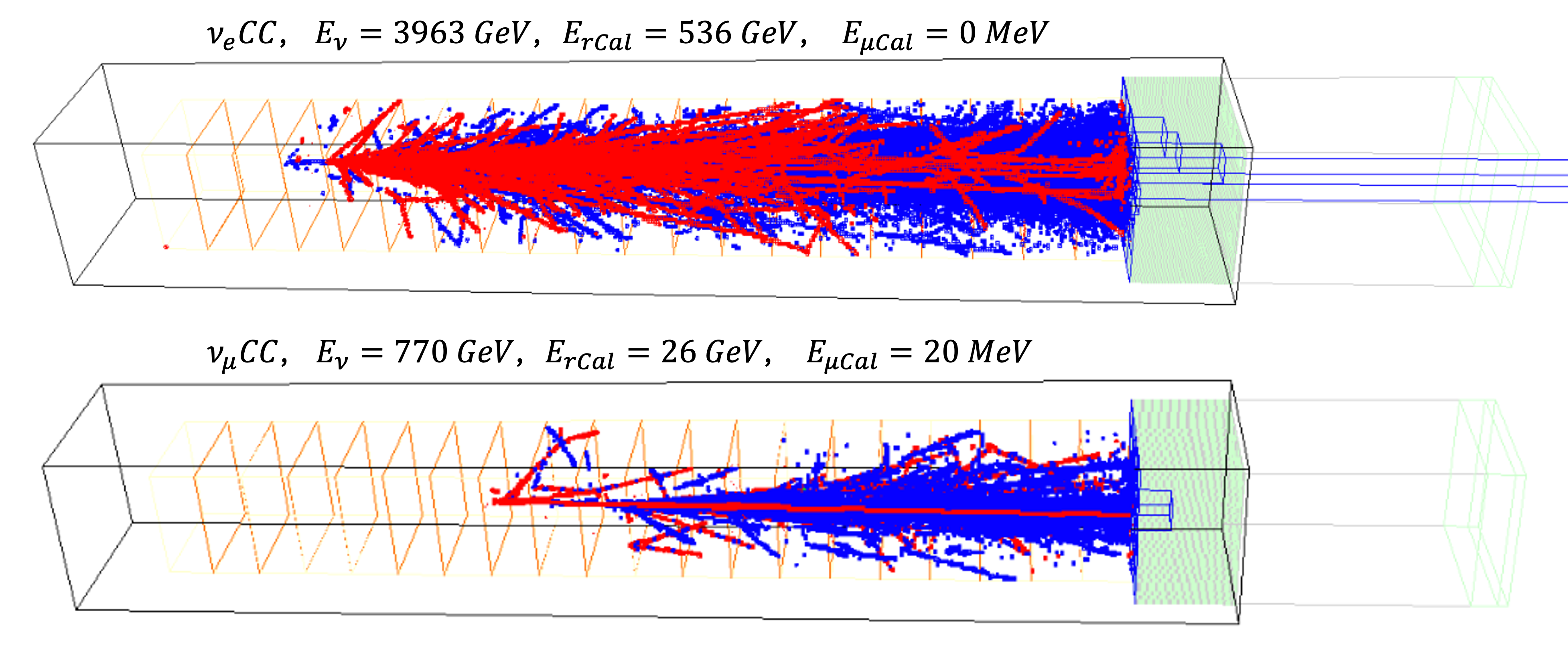}
\caption{Simulated $\nu_e$ and $\nu_\mu$ CC interactions in the FASERCal detector. The electromagnetic shower component is shown in blue, while the hadronic component is displayed in red. } \label{fig:FASERCAL_nuevent}
\vspace{-3mm}
\end{wrapfigure}
This design is to be combined with the high-precision tracking detector. Simulated neutrino interactions are shown in Figure \ref{fig:FASERCAL_nuevent}.
The picture highlights the ability of the highly segmented scintillator voxel detector to accurately track and reconstruct both electromagnetic shower and muon-induced signals, ensuring high precision in the identification and measurement of neutrino interactions at TeV energy levels. 
The possibility of reconstructing the energy flow in 3D opens the door to measure exclusive final states and exploit the transverse kinematical variables with the possibility of using these features to identify charm and tau events.\medskip 

\section{Upgraded emulsion detector for Run~4\label{sec:emulsionsrun4}}

\begin{wrapfigure}{r}{0.6\textwidth}
\vspace*{-3mm}
\includegraphics[width=0.6\textwidth]{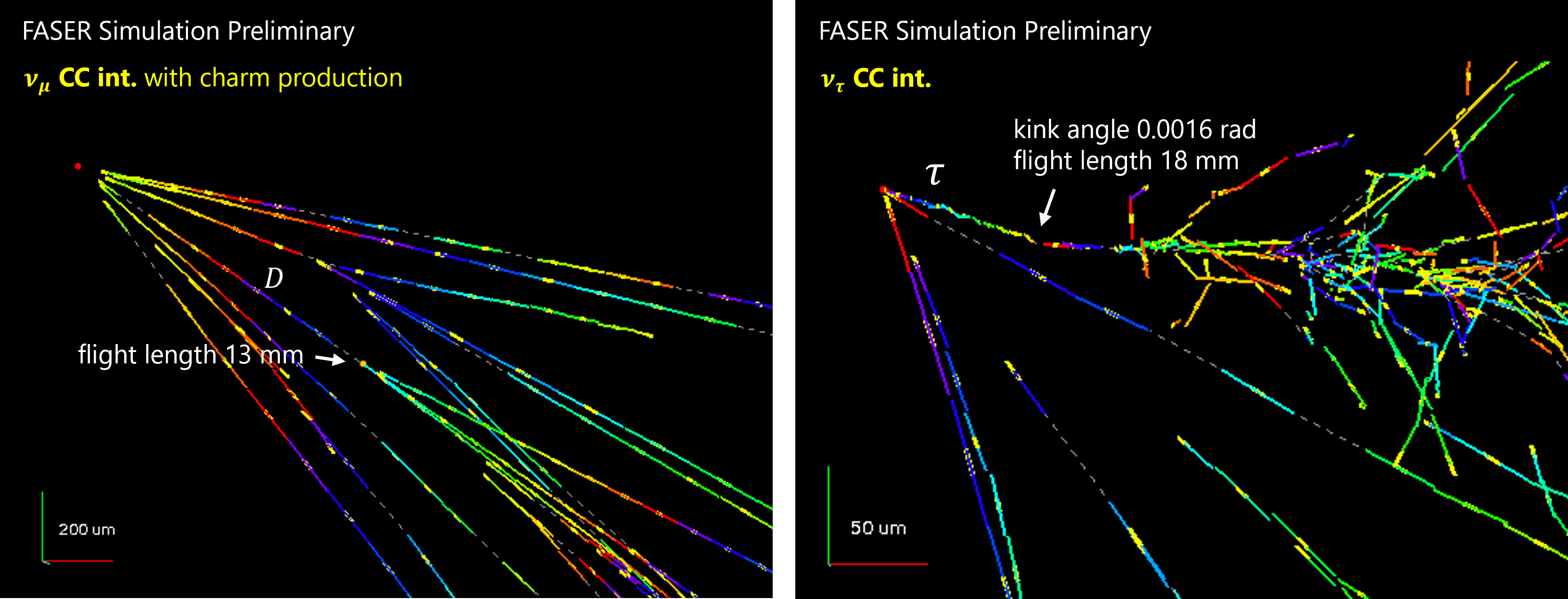}
\caption{Simulated charm production and $\nu_\tau$ CC interactions in emulsions. } \label{fig:fasernu_tau}
\vspace*{-3mm}
\end{wrapfigure}

The photographic emulsion technique~\cite{Ariga:2020lbq} enables three-dimensional track reconstruction with sub-micrometer-level resolution (see \cref{fig:fasernu_tau}). There is extensive expertise in detecting neutrino interactions using emulsions, and the current FASER$\nu$ experiment has already demonstrated its successful application in studying LHC neutrinos,~\cite{FASER:2021mtu, FASER:2024hoe} in particular at detecting electron neutrinos and providing for the first time measurements of cross-sections. However, since emulsions integrate over time, they are sensitive to the total fluence of muons, and increasing the frequency of replacing the emulsion stack more often than in Run 3 is not feasible due to the limited number of permitted accesses to the underground area, which interfere with LHC operations.
Therefore, ongoing studies are exploring whether an upgraded emulsion program could be implemented by leveraging improved technologies to process emulsions and reconstruct neutrino interactions with high efficiency, even in a high-background environment. In parallel, an implementation of sweeping magnets upstream to reduce the muon background is under study. 
An enhanced emulsion detector for Run 4 would provide the best solution for studying $\nu_\tau$ and $\bar \nu_\tau$ CC interactions on an event-by-event basis with high specificity,
enabling the test of lepton universality among all three neutrino flavors. It would also allow high-statistics measurements of exclusive charm production in high-energy neutrino interactions. In the case that the emulsion detector would not be able to cope with the full data-taking period, the beam time may be shared with other electronic detectors, presumably at the beginning or the last period of each year.

\vspace{-3mm}
\section{Project Planning \label{sec:project}}

The current plan is to implement the upgrade of the FASER detector by the start of the LHC Run 4.
Several FASER institutions have started an active R\&D program to define possible detector designs.
We are confident about the timescale. The FASER collaboration, while relatively small, is fully established, very functional, and fully adapted to tackle the challenge of the upgrade. 
The collaboration also already has extensive experience efficiently installing detectors in the TI12 area
A clear advantage of our proposals is that no civil engineering nor significant modifications in the TI12 tunnel are required.
The foreseen cost is of the order of a few million CHF, which is affordable and totally realistic given the size of the Collaboration.

\vspace{-3mm}
\section{Final considerations}

High-statistics TeV neutrinos at the LHC, enabled by its high luminosity, provide valuable insights into the highest man-made energies, which are otherwise inaccessible. An upgraded FASER detector presents a practical and cost-effective solution to seize this opportunity, beginning with LHC Run 4 and extending beyond.

The FASER experiment in Run 4 serves as a platform to explore new detector technologies for optimally studying TeV-scale neutrinos and generating high-statistics data for physics analyses. It is also crucial for maintaining expertise in neutrino physics within the research community.  It opens the path towards a potential program at the Forward Physics Facility~\cite{Anchordoqui:2021ghd, Feng:2022inv, Adhikary:2024nlv}, where detectors with 20-ton scale targets could take maximal advantage of the LHC neutrino beam with unprecedented event rates for high precision measurements and stringent beyond SM searches.

FASER operates on a human scale, making it particularly well-suited for early-career researchers. Its timeline aligns perfectly with a graduate student’s career, allowing them to engage in detector construction, commissioning, and data analysis within a single PhD program. The experiment also offers multiple leadership opportunities for young researchers. Additionally, FASER is highly sustainable, as the neutrino beam is readily available whenever the LHC is operational.

Overall, FASER is a compact and fast-paced experiment, easy and cheap to operate, that enables hands-on experience in detector development, commissioning, and data analysis.
It represents a unique opportunity to perform neutrino physics in Europe not reachable elsewhere, attracting researchers worldwide.

\clearpage
\section*{Acknowledgments}
\label{sec:Acknowledgments}


The FASER Collaboration gratefully thanks the technical and administrative staff at CERN as well as other FASER institutions for their invaluable contributions to the success of the FASER experiment. We also thank the CERN Physics Beyond Colliders study group for important support as well as Hiroaki Menjo and Kazuhiro Watanabe for many useful discussions. We further thank Jianming Bian, Mingyi Dong, Shu Li, Jianbei Liu, Yong Liu, Wataru Ootani, Zhongtao Shen, Michael Smy, Konstantinos Spyrou, Takeshita Tohru, Zhigang Wang, Wenjie Wu, Haijun Yang, Boxiang Yu, and Yunlong Zhang for contributions to the presented upgraded detector proposals.

This work was supported in part by Heising-Simons Foundation Grant Nos.~2018-1135, 2019-1179, and 2020-1840, Simons Foundation Grant No.~623683, U.S. National Science Foundation Grant Nos.~PHY-2111427, PHY-2110929, and PHY-2110648, JSPS KAKENHI Grant Nos.~19H01909, 22H01233, 20K23373, 23H00103, 20H01919, and 21H00082, the joint research program of the Institute of Materials and Systems for Sustainability, ERC Consolidator Grant No.~101002690, BMBF Grant No.~05H20PDRC1, DFG EXC 2121 Quantum Universe Grant No.~390833306, Royal Society Grant No.~URF$\backslash$R1$\backslash$201519, UK Science and Technology Funding Councils Grant No.~ST/ T505870/1, the National Natural Science Foundation of China, Tsinghua University Initiative Scientific Research Program, and the Swiss National Science Foundation.

\bibliographystyle{utphys}
\bibliography{references}

\end{document}